\documentclass[symmetry,article,accept,pdftex,moreauthors]{Definitions/mdpi}

\firstpage{1} 
\pubvolume{1}
\issuenum{1}
\articlenumber{0}
\pubyear{2024}
\copyrightyear{2024}
\externaleditor{\textls[-25]{Academic Editor: Andrea Lavagno}}
\datereceived{11 November 2024} 
\daterevised{12 December 2024} 
\dateaccepted{16 December 2024} 
\datepublished{ } 
\hreflink{\textls[-25]{https://doi.org/}} 

\usepackage{amsmath}
\usepackage{amssymb}
\usepackage{nccmath}
\usepackage{bm}
\usepackage{float}
\usepackage{slashed}
\usepackage{rotating}
\usepackage{braket}
\usepackage{placeins}
\usepackage[labelformat=simple]{subcaption}

\DeclareCaptionLabelFormat{subcaptionlabel}{\normalfont(\textbf{#2}\normalfont)}
\Title{Monopole and Seniority Truncations in the Large-Scale Configuration Interaction Shell Model Approach}

\TitleCitation{Monopole and Seniority Truncations in the Large-Scale Configuration Interaction Shell Model Approach}

\Author{{Priyanka Choudhary * and Chong Qi *} 
}

\AuthorNames{Priyanka Choudhary and Chong Qi}

\AuthorCitation{{Choudhary}
, P.; Qi, C.}

\address[1]{Department of Physics, KTH Royal Institute of Technology, 10691 Stockholm, Sweden}

\corres{\hangafter=1 \hangindent=1.05em \hspace{-0.82em}Correspondence: pricho@kth.se (P.C.); chongq@kth.se (C.Q.)}

\abstract{This paper addresses the challenges of solving the quantum many-body problem, particularly within nuclear physics, through the configuration interaction (CI) method. Large-scale shell model calculations often become computationally infeasible for systems with a large number of valence particles, requiring truncation techniques. We propose truncation methods for the nuclear shell model, in which angular momentum is conserved and rotational symmetry is restored. We introduce the monopole-interaction-based truncation and seniority truncation strategies, designed to reduce the dimension of the calculations. These truncations can be established by considering certain partitions based on their importance and selecting physically meaningful states. We examine these truncations for Sn, Xe, and Pb isotopes, demonstrating their effectiveness in overcoming computational limits. These truncations work well for systems with either a single type of valence nucleon or with both types. With these truncations, we are able to achieve good convergence for the energy at a very small portion of the total dimension.}


\begin{document}

\section{Introduction}
The solving of the quantum many-body problem is essential for the understanding of various microscopic systems like nuclei, atoms, and condensed matter, as well as quantum chemical multi-electron systems. In~the present super-computation and machine-learning era, the~importance of ab~initio modelling has been repeatedly emphasized. Of~all the ab~initio many-body methods~\cite{HERGERT2016165,PhysRevC.93.051301,PhysRevC.96.034324,PhysRevC.105.034333,PhysRevC.64.014001,PhysRevC.76.064319,PhysRevC.78.065501,PhysRevLett.113.142502,PhysRevC.82.034330,Hagen2016,BARRETT2013131,PhysRevLett.106.192501} that have been developed, the~configuration interaction (CI) approach~\cite{doi:10.1080/002689798168303,brussaard1977shell,Caurier2005427,annurev:/content/journals/10.1146/annurev.ns.38.120188.000333,Tsunoda2020,heyde1998nucleons} may be one of the easiest to understand and most convenient to use but it is also one of the most difficult tools to implement efficiently on a supercomputer. CI involves essentially a matrix-eigenvalue equation involving an atomic or nuclear many-body Hamiltonian. One can expect to attain the exact solution of the many-body Hamiltonian if the full configuration interaction calculation involving all possible orbital bases can be carried out, which, however, is often not the case. As~the dimension of the CI problem increases exponentially as the number of particles increases, in~most cases one has to implement various truncation algorithms. By~truncation, one aims at limiting the number of basis states for the CI model space. Unlike coupled-cluster~\cite{PhysRevLett.113.142502,PhysRevC.82.034330,Hagen2016} or many-body perturbation theory~\cite{ROTH2010272,TICHAI2018195} approaches, the~approximate solution of the CI after truncation can still be deemed as variational. One may be able to recover different parts of the correlation energy depending on the truncation strategy applied. Other CI approaches like the Monte Carlo shell model~\cite{PhysRevLett.75.1284,PhysRevLett.81.1588,physics4030071,PhysRevC.98.054309}, variation after projection~\cite{GAO2022136795}, and generator coordinate method~\cite{sym16040409} have also been developed to overcome the dimension difficulty, but they often started from deformed rather than spherical Slater determinants and  therefore require very different~algorithms. 

Another important aspect to consider when designing a truncation algorithm is the possible breaking of the underlying symmetry. The~restoration of the symmetry may be quite computationally heavy. In~this paper, we focus in particular on the nuclear many-body system where the angular momentum is conserved and therefore the rotational symmetry is important to restore. We will review very briefly the different truncation algorithms that have been applied in nuclear CI approaches. Then we will introduce the monopole-interaction-based importance truncation (referred to as monopole truncation) and further truncation by considering the seniority truncation as inspired by the presence of strong pairing correlation. Finally, we will test these truncations on the system that contains only one type of valence nucleon, with~an equal number of valence protons and neutrons and~with an unequal number of valence protons and~neutrons. 

The paper is divided into the following sections: In Section~\ref{SecII}, we have briefly described the configuration interaction shell model and different shell model algorithms, focusing on the angular momentum projection in the NushellX formalism. We have proposed different truncation approaches in Section~\ref{SecIV} in order to overcome the dimensionality issue. In~Section~\ref{SecV}, details are given about the model spaces and the effective interactions used in the present calculations. Shell model results of low-lying energy spectra with full and truncated basis states for tin, lead, and~xenon isotopes are presented in Section~\ref{SecVI}. We have also calculated reduced electric quadrupole transition probability $B(E2)$ for Sn and Xe isotopes. Finally, conclusions are drawn in Section~\ref{SecVII}.

\section{The Nuclear Configuration Interaction Shell~Model}\label{SecII}
The CI approach is more commonly referred to as the nuclear shell model in the nuclear physics community, as it has its root in the independent particle model (the ``shell model'') that was introduced 75 years ago) \cite{PhysRev.74.235,PhysRev.75.1969,PhysRev.75.1766.2,PhysRev.78.16}.
The nuclear shell model has long been one of the most successful and accurate approaches in describing the nuclear structure properties. To~put the nuclear shell model in a general context, one may take the independent particle model as the starting point, which may be well estimated from Hartree-Fock (HF) approaches, and~the CI with effective interactions as the post-HF treatment. CI calculations can be carried out without or with assuming an inert core. The~latter is more common, as the computation is often too heavy to include the core excitations and by considering the fact that the atomic nucleus is indeed characterized by strong shell~effects.

A nucleus is considered to be made of interacting $A$-nucleons (with $Z$ protons and $N$ neutrons) outside a frozen inert core. The~Hamiltonian for such a system of $A$-nucleons can be expressed in terms of single-particle energies and two-body matrix elements such as
\begin{equation}
    H_A = T + V = \sum_{\alpha}{\epsilon}_{\alpha} a^{\dagger}_{\alpha} a_{\alpha} + \frac{1}{4} \sum_{\alpha\beta\delta\gamma JT} \langle j_{\alpha} j_{\beta} | V | j_{\gamma} j_{\delta} \rangle _{JT} A^{\dagger}_{JT:{j_{\alpha}}{j_{\beta}}} A_{JT:{j_{\delta}}{j_{\gamma}}}.
\end{equation}
where $\alpha$ defines a single-particle state within a given model space with the set of quantum numbers $\{n,l,j,t\}$ and the single particle energy corresponding to the state is given by $\epsilon_{\alpha}$. The~operators $a^{\dagger}_{\alpha}$ and $a_{\alpha}$ are the creation and annihilation operators, respectively. $\langle j_{\alpha} j_{\beta} | V | j_{\gamma} j_{\delta} \rangle _{JT}$ are the antisymmetrized two-body matrix elements, and $A^{\dagger}_{JT}$ and $A_{JT}$ are the fermion pair creation and annihilation operators, respectively. With~the above-expressed effective Hamiltonian, the~eigenvalue and eigenfunctions can be calculated in the full configuration interaction shell model approach. 
The corresponding shell model energy of the state can be given by
\begin{equation}
    E_{\rm SM}^{\rm Cal.} = \bra{\Psi} H_A \ket{\Psi}
\end{equation}
where $\Psi$ is the calculated shell-model wave function of the state. The~ground state energy can be calculated by variational methods as well, which is equivalent to the CI~approach.

In practice, one commonly applies the frozen-core approximation and limits the single-particle orbitals to be the few that are just above the core orbitals (usually those within one or a few major shells). Those orbitals define the model space, while all orbitals above the model space are skipped.
The single-particle energies can be taken from experimental data, HF calculations, and/or fitting to experimental data. The~two-body matrix elements can be evaluated from realistic nucleon-nucleon potentials via standard perturbation theory or in-medium Similarity Renormalization Group methods~\cite{KUO196640,PhysRevC.65.061301} and by fitting to experimental data~\cite{COHEN19651,PhysRevC.74.034315,PhysRevC.85.064324}. Very often the effective Hamiltonian is provided as a list of numbers, including the single-particle energies and two-body matrix elements for a given model space. Most of the existing CI codes are designed in that way.
We ignore here the three-body matrix elements. There have been recent efforts in including the three-body effects either at the mean-field (single-particle) or in explicit three-body matrix elements~\cite{COLE197524,ANDREOZZI1976388}. The~latter is straightforward to implement in existing CI algorithms but the inclusion of three-body matrix elements can make the Hamiltonian matrix very dense and hard to handle on a~supercomputer.

There are several efficient shell model codes available to construct the basis states and the Hamiltonian matrix and then solve the Hamiltonian matrix with the Lanczos~\cite{Lanczos:1950zz} or similar iterative methods. Those codes are based on either angular momentum coupled J(T)-scheme (with parity, spin, and optionally isospin conserved) or uncoupled $M$-scheme (with only parity and total $M$ quantum number conserved). The~$M$-scheme is the de facto standard choice for large-scale calculations, as it is easier to handle on a modern computer. There are public codes in the $M$-scheme available, like BigStick~\cite{JOHNSON20132761} and KSHELL~\cite{SHIMIZU2019372}, both of which are highly optimized for parallel calculations on supercomputers with hybrid MPI+openMP algorithms. An~$M$-scheme calculation can be implemented efficiently starting with
basic bit operations where each basis (Slater determinant) can be represented as an integer. 
The disadvantage of the $M$-scheme is that the dimension maximizes. The~dimension for a problem with certain $J$ instead of $M$ in the angular-momentum-coupled bases can be much smaller. The~angular momentum coupling can be carried out via the coefficients of fractional
parentage (cfp) \cite{book:91951242}, in~multisteps~\cite{XU201251}, or through the correlated basis method re-using part of the rotationally-invariant Hamitonian as the angular momentum projector~\cite{PhysRevC.90.024306}.

\subsection{$M$-Scheme~Algorithms}

The many-body Hamiltonian is invariant under rotations, which means that the total $\hat{J^2}$ and $\hat{J_z}$ operators commute with the Hamiltonian. Consequently, the~eigenstates of the Hamiltonian have both $J$ and $M$ as good quantum numbers. Because~the Hamiltonian cannot connect many-body states with different 
$M$, it becomes advantageous to use the so-called $M$-scheme basis, where all many-body basis states share the same {$M$} value. This approach is particularly convenient since $M$ is an additive quantum number. To~determine the M for a Slater determinant, one can add the $m_i$ of the occupied single-particle states. The~eigenvalue can only depend upon $J$, not $M$, allowing the $M$-scheme to eliminate the rotational degeneracy and significantly reduce the size of the basis. When dealing with systems involving two types of particles, the~structure of the many-body basis becomes more complicated. In~such cases, the~concept of factorizations is used to represent the many-body basis in order to manage this~complexity.

In the case of a nucleus, we have two species of fermions, protons and neutrons. Any wavefunction can be expanded as a sum of product wavefunctions of protons and neutrons.
\begin{equation}
    \ket{\Psi} = \sum_{\mu \nu} c_{\mu \nu} \ket{\mu_{p}} \ket{\nu_n}
\end{equation}

One constructs the many-body state $\ket{\mu_{p}},  \ket{\nu_n}$ from a finite set of orthonormal single-particle states ${\phi_i}$. The~single-particle states must have a good quantum number as total angular momentum, $j$, and $z$-component of the angular momentum, $m$, and~parity $\pi$.
One can think of the single-particle states as eigenstates of a rotationally invariant single-particle Hamiltonian. For~a given state ${\phi_i}$, we need to know $j_i$, $m_i$, and $\pi_i$, where all possible $m_i$ are allowed for a given $j_i$. 
We construct the many-body states from the single-particle states as Slater determinants using the antisymmetrized product of single-particle states. 

On computers, it is convenient to represent the occupation of single-particle states using bit representations, with~occupied states represented by bit 1 and unoccupied states by bit 0. This factorization makes the calculation of the may-body matrix elements of Hamiltonian simplified and the number of non-zero matrix elements is limited. Another \mbox{$M$-scheme}-based shell model code is KSHELL developed by N. Shimizu and \mbox{collaborators~\cite{SHIMIZU2019372}.}

\subsection{Angular Momentum Projection and the NuShellX~Code}\label{2.2}
The NuShellX code is one of the most popular CI codes, which gives exact eigenenergies and eigenwave functions in angular momentum coupled bases. Here, the~code starts with a $M$-scheme basis, and the angular momentum is restored by applying a projection algorithm~\cite{RevModPhys.36.966}. That hybrid algorithm was also applied in older OXBASH code~\cite{Langanke:1991gox} and our in-house code~\cite{qi2007modernshellmodeldiagonalizationsrealistic}. 

The starting point for such an algorithm is to generate a set of bases
with a good magnetic quantum number in the $M$-scheme. Neutrons and protons
can be naturally blocked into two spaces. The~shell model codes {ANTOINE and NATHAN} \cite{etde_359035} proposed the idea of factorization into complementary parts, which reduces the requirement of storing the many-body Hamiltonian matrix. The~dimensions in the proton and neutron spaces separately are small, even for large dimensions in the total space. The~bases are classified according to
the distributions of particles in the single-particle orbits
(referred to as partitions). Since the angular-momentum projection operator can only
change the magnetic quantum number, the~projection can be completed for
each partition separately. Within~each partition,
vectors with good angular momentum can be expanded in $M$-scheme bases as
\begin{equation}
	\label{Eq:4}
|\Psi^{J}_i\rangle=\sum_{m\leq i}
\mathcal{M}_{im}P^{J}|\alpha_m\rangle,
\end{equation}
where $P^J$ is the projection operator and $|\alpha\rangle$ a set of
specially chosen $M$-scheme bases. $\mathcal{M}$ is a lower triangle
matrix. Vectors with good angular momentum are the orthonormal \linebreak and
satisfy,
\begin{eqnarray}
\nonumber
\langle\Psi^{J}_i|\Psi^{J}_j\rangle&=&\langle\Psi^{J}_i|\sum_{m\leq
j} \mathcal{M}_{jm}P^{J}|\alpha_m\rangle\\
\nonumber &=&\sum_{m\leq j} \mathcal{M}_{jm}\langle
\Psi^{J}_i|\alpha_m\rangle\\
&=&\delta_{ij}.
\end{eqnarray}
Elements of $\mathcal{M}$ can be obtained by inverting the matrix
$\langle \Psi^{J}_i|\alpha_m\rangle$. Since the projection operator
satisfies $P^J\cdot P^J=P^J$, the~$n$-th vector can be obtained
through,
\begin{equation}
|Q_n\rangle=|\alpha_n\rangle-\sum_{i<n}\langle O_i|\alpha_n\rangle
\sum_{j\leq i}\mathcal{M}_{ij}|\alpha_j\rangle,
\end{equation}
and
\begin{equation}
|\Psi^J_n\rangle=P^J|Q_n\rangle(\langle Q_n|P^J|Q_n\rangle)^{-1/2}.
\end{equation}
For the angular momentum projection, one has the freedom to restore the total angular momentum or to do the projection separately for the proton and neutron blocks and couple them to certain total angular momentum (via standard angular momentum coupling).

The angular momentum projection can still be quite time-consuming, but it is much easier than standard cfp calculations. The~advantage is that, after~the projection, the~dimension of the coupled bases would be much smaller, which makes it easier for the diagonalization.
The disadvantage is that one has to store a large amount of angular-momentum projection coefficients either in memory~\cite{qi2007modernshellmodeldiagonalizationsrealistic} or on a disk as in the case of NuShellX~\cite{Nushellx}. 
NuShellX was written in Fortran90 by W.D.M. Rae. The~proton and neutron angular momentum projections are made separately using the NuShell code, which are coupled together to make the total wave function. The~NuShell is the modified version (more accurate and significantly faster) of the original $JT$ projection code OXBASH~\cite{Nushellx}. In~the version we have been developing at KTH, the~code is separated into different independent codes including NuBasis (which generates the list of partitions and the $M$-scheme bases), NuProj which implements the angular momentum projection and gives the $J$-scheme bases defined in the $M$-scheme bases, NuMatrix which generates the pp and nn matrices in a semi-orthogonal basis, NuOrth that completes the orthogonalization of the $J$-scheme matrix produced by NuMatrix, NuOper (converts the interaction into an m-scheme operator), NuOp produces a $J$-scheme particle-hole transformation of the np interaction, NuOpm calculates all the $m$-scheme operators, NuOpd converts the $m$-scheme operator matrix elements to reduced matrix elements in $J$-scheme, NuLnz (the standard Lanczos module), and~NuVec that gives the eigenvectors and~eigenvalues.    

\section{Truncation}\label{SecIV}
The shell model calculations for heavy isotopes with a large number of valence particles can easily go beyond the capability of the most advanced computational resources due to the huge $M$-scheme dimension. A~truncation criterion must be implemented in most calculations. Several works have been carried out by Horoi and co-workers~\cite{PhysRevC.50.R2274,PhysRevLett.82.2064,PhysRevC.65.027303} in order to  develop the exponential convergence method (ECM), which reduces the whole configuration space into a subspace by concerning the average centroid of partitions. A~simple and natural approach is to restrict the number of particle-hole excitations across a major shell gap or a subshell closure. It is called n-particle-n-hole truncation. If~there is more than one major shell included in the model space, $n\hbar\omega$ truncation can be applied to limit the number of excitations crossing the major shells (which is often the case in no-core shell model calculations). Such truncation is easy to implement in both $M$-scheme and $J$-scheme codes, which essentially only limits the number of partitions in the calculation. All $M$-scheme or $J$-scheme bases within the kept partition are included, which avoids the symmetry break effect. Such truncation may not be very effective.
Instead, in~such cases, and if there is no major-shell or subshell closure, we have introduced monopole-based truncation in the shell model diagonalization method, which is described below in detail. In~the present work, we have implemented the truncations in the KTH version of the popular shell model code NuShellX (where the code sources and inputs are simplified) \cite{Nushellx}.

\textls[-10]{One may design an algorithm that directly truncates the $M$-scheme bases. Methods like density matrix renormalization group~\cite{PhysRevC.65.054319,PhysRevC.73.014301,PhysRevLett.97.110603,TICHAI2023138139} and importance \mbox{truncation~\cite{Andreozzi_2003,PhysRevC.93.021301,Bianco_2010,PhysRevC.93.064328}.} We have also introduced a truncation based on a pseudo-seniority-like truncation (C. Qi and N. Shimizu (unpublished).) in \mbox{$M$-scheme~\cite{Qi849105}.}  The truncation in the $M$-scheme may, however, lead to the breaking of the rotational symmetry and the angular momentum conservation and, as a result, \mbox{convergence~issues.}}

\subsection{Monopole-Based~Truncation}
The single-particle energy in the Hamiltonian can be deemed as the HF state energy. In~principle, for~a system with a large number of valence particles, one may consider starting with an HF calculation for the present shell-model Hamiltonian and re-express the two-body matrix elements in the new HF bases. There are methods developed in that or similar manners. However, it can be difficult to implement numerically, as the HF bases thus calculated may very well be deformed. Instead,
the total Hamiltonian can be re-written as~\cite{POVES1981235}
\begin{equation}
	H_A = H_{m} + H_{M}.
\end{equation}
Where $H_m$ and $H_M$ represent the monopole and multipole Hamiltonians. The~energy corresponding to the monopole Hamiltonian is expressed as
\begin{equation}
	E_m = \bra{\Psi}H_{m}\ket{\Psi} = \sum_{\alpha} \epsilon_{\alpha} \braket{\hat{N}_{\alpha}} + \sum_{\alpha \leq \beta} V_{m;\alpha\beta} \braket{\frac{\hat{N}_{\alpha}(\hat{N}_{\beta} - \delta_{\alpha\beta})}{1+\delta_{\alpha\beta}}}.
\end{equation}
{In} the above expression, $\sum_{\alpha} \braket{\hat{N}_{\alpha}} = N$ the total number of valence particles and $$\sum_{\alpha \leq \beta} \braket{\frac{\hat{N}_{\alpha}(\hat{N}_{\beta} - \delta_{\alpha\beta})}{1+\delta_{\alpha\beta}}} = \frac{N(N-1)}{2}.$$

The monopole interaction $V_{m;\alpha\beta}$ is expressed as the angular momentum weighted average value of the diagonal matrix elements for a given set of $j_{\alpha}$, $j_{\beta}$, and~$T$.
\begin{equation}
	V_{m;\alpha\beta} = \frac{\sum_J (2J + 1)\bra{j_{\alpha}j_{\beta}}V\ket{j_{\alpha}j_{\beta}}_J}{\sum_J (2J + 1)}.
\end{equation}

The monopole interaction, together with the single-particle energies, determines the mean field/bulk properties of the shell-model Hamiltonian, while the residual multipole interactions are essential for the mixture of different Slater determinants and the \linebreak correlation~energy. 

The wave function is constructed as a linear combination of all possible antisymmetric Slater determinants within a valence space. The~valence space, or model space, is a set of single-particle orbitals near the Fermi surface. In~the full configuration interaction shell model calculations, one needs to define the basis in terms of {$partitions$}. A {$partition$} is defined as a set of configurations with the same number of particles in each single-particle orbital. The~total number of partitions for Pb isotopes with valence neutrons are presented in Ref.~\cite{PhysRevC.94.014312}.
Then, the~wave function is constructed for each partition in the $j-j$ coupled scheme or the uncoupled $M$ scheme. 

In the $M$-scheme, $M (j_z)$ and $T_z$ are good quantum numbers only; angular momentum is not explicitly. It is difficult to implement the truncations, which leads to large dimensions of the bases. If~we remove part of the bases within a given partition, it will create problems. Instead, it will be convenient if one includes only a limited number of partitions and considers all the $M$-scheme bases within a given partition. We can simply include some partitions and exclude the rest to reduce the dimension, according to their importance.  
We implemented an importance truncation based on the total monopole energy by taking the multipole Hamiltonian as a perturbation. The~monopole energy for a given partition is written as
\begin{equation}\label{Eq.11}
  E_m^P = \bra{\Psi}H_{m}\ket{\Psi} = \sum_{\alpha} \epsilon_{\alpha} N_{P;\alpha} + \sum_{\alpha \leq \beta} V_{m;\alpha\beta} {\frac{N_{P;\alpha}(N_{P;\beta} - \delta_{\alpha\beta})}{1+\delta_{\alpha\beta}}},
\end{equation}
where $N_{P;\alpha}$ is the distribution of the valence particles within a given partition $P$. First, one needs to calculate the monopole energies for all partitions (which is carried out using \mbox{Equation~(\ref{Eq.11})}), and all $M$-scheme basis states for a given partition have the same monopole energy. One needs to determine the minimum monopole energy. We defined the cutoff energy $E_{\rm cutoff}$, which is the energy relative to the lowest one of all the partitions. We select the partitions whose energies are smaller than the cutoff energy and create a subspace in which the wavefunction is spanned. This truncation is referred to as ``sharp cutoff truncation.'' If~we set the $E_{\rm cutoff}$ equal to its maximum value, all the partitions are taken into account, and~the dimension corresponds to the full space dimension. As~the cutoff energy increases, the~number of configurations (partitions) included also increases, and~consequently, the~dimension of the subspace increases. This implies that by removing restrictions, we approach the full space~calculation. 

Apart from the sharp cutoff truncation, we propose an idea of the ``distribution type cutoff truncation'' for the first time. We consider a distribution function that is defined as
$$ X = \frac{1}{ 1+e^((E_{\rm mono}-E_{\rm mono\textunderscore min}-E_{\rm cutoff})/a)},$$
where $E_{\rm mono}$ is the monopole energy for a particular partition, $E_{\rm mono\textunderscore min}$ and $E_{\rm cutoff}$ represent the minimum monopole energy and the sharp cutoff energy, respectively, and a defines the smoothness of the distribution curve. We use a random number generator to select the bases in such a way that if the generated random number for a partition is less than the defined function $X$, the~corresponding basis states are included; otherwise, they are excluded. The~idea behind the distribution-type cutoff is to include important partitions---those excluded in the sharp cutoff monopole truncation---based on the smoothness of the~function.

In the third case, we remove a part of the $J$-scheme bases associated with each partition based on function $X$. We define the allowed basis as $X$ times the number of total bases ($X * nJT$). To~check these different types of monopole-based truncations, we have performed the shell-model calculations for various isotopes. There have been attempts in the past to use an approach similar to the monopole-based truncation, and this is based on spectral distribution theory, described in Refs.~\cite{Wong1,Wong2,Wong3}.

\subsection{Seniority~Truncation}
The seniority scheme provides a good approximation for the low-lying states of systems containing the same kind of particles~\cite{talmi1993simple}. This arises from the fact that monopole pairing interactions with $J = 0$ dominate the $T = 1$ two-body matrix elements. The~seniority quantum number is defined as the number of unpaired nucleons in a nuclear state, i.e.,~the number of particles not coupled to $J = 0$. Recent works on the study of the seniority coupling scheme are reported in Refs.~\cite{PhysRevC.82.014304,PhysRevC.83.014307,QI201221}. In~Ref.~\cite{XU2013247}, it is reported that the pairing Hamiltonian, described by the $v = 0, J = 0$ states in many shells, can provide the exact solution. These states show only a small fraction of the total wave function but represent the most significant components for describing low-lying nuclear states. Interestingly, the~number of $v = 0$ states is even smaller than the total number of partitions. Alternatively, $v = 0$ states can be constructed within the $M$-scheme, which offers a straightforward approach. However, the~dimension of the bases in such a scheme is considerably larger compared to that in the $jj$-scheme.

It is not trivial to define seniority in an uncoupled $M$-scheme or partition. Instead, we start simply with a seniority-like or quasi-seniority scheme in the $M$-scheme instead and define a nucleon pair as seniority (or quasi-seniority)-0 if they couple to the $M$ quantum number $M=0$ (of course, in~practice such pairs can exhibit non-zero angular momentum and be of seniority two, strictly speaking). We hope that such configurations are the most favored components of the low-lying states, as they are expected to be dominated by low-seniority states that can be projected from those pairs.
In NuShellX, a~projection algorithm~\cite{RevModPhys.36.966} is applied to store the angular momentum as described in Section~\ref{2.2}. The~angular momentum projection method~\cite{RevModPhys.36.966} starts with a random $M$-scheme basis $|\alpha_m\rangle$ of Equation~(\ref{Eq:4}) on which we make an important selection. A~random selection of $M$-scheme basis leads to a random set of $J$-states, which means that the eigenvectors obtained may correspond to different $J$ values. Instead, a~more effective and meaningful strategy is to begin with the basis that already has a good angular momentum $J$. This ensures that all the eigenvectors thus obtained have the same angular momentum. The~chosen basis state should contain the most significant features about the Hamiltonian within each partition. Compared to the random projected basis used in the projection approach~\cite{RevModPhys.36.966}, these bases are more physically meaningful. One
thus has the opportunity to perform different truncation schemes within this approach. Basis states with lower seniority are expected to play a crucial role in describing low-lying states. Ground states of even-even nuclei are generally assumed to have zero seniority (fully paired), whereas those of odd-$A$ nuclei typically have seniority one. We implement a truncation to the selection of $M$ states with low seniority, so that we end up with the $J$-states that are relevant to the ground state. We introduce the seniority quantum number ($\nu$) as $\nu = N - 2* Np,$ where $N$ is the number of valence nucleons and~$Np$ the number of paired nucleons. Firstly, we check the number of nucleon pairs and determine the seniority of all $M$-scheme bases within each partition. For example, the~$M$-scheme bases are 1111011 and 10001011011 in $^{202}$Pb. These states have 3 and 2 pairs, respectively. Thus the respective seniority quantum numbers are 0 and 2. Then, we identify the minimum seniority for each partition. Finally, we do the projection by selecting a starting $M$-state with low~seniority.

Pairing correlations are anticipated to play a crucial role in describing the lowest-lying states. Generalized seniority truncation has been proposed as a truncation scheme for large-scale shell model calculations~\cite{talmi1993simple,Jia_2015}. This approach has been applied to the Sn isotopes~\cite{PhysRevC.94.044312,MAHESHWARI201662} and Pb isotopes, considering states with generalized seniority up to S = 6. The~nucleon-pair approximation (NPA) is a pair-truncated shell-model approach with collective pairs as building blocks, which has been used in Refs.~\cite{PhysRevC.88.044332,PhysRevC.89.014320} for the shell model calculations. In~Ref.~\cite{Qi849105}, various truncation schemes for the nuclear configuration interaction shell model approach have been reported. Our method is more general, and it selects the most important partitions determined by the monopole interaction and starts with low-seniority states within that~partition.

\section{Model Space and the Effective~Interaction}\label{SecV}
We have selected tin, xenon, and lead isotopes for the shell model~study. 

\subsection{Lead~Isotopes}

For the detailed structural description of lead isotopes, we have used the interaction developed by the Stockholm group (J. Blomqvist and C. Qi (unpublished).)~\cite{PhysRevC.94.014312}. For~the interaction, doubly magic $^{208}$Pb is assumed as the inert core. The~model space is made up of six orbitals: 2$p_{1/2}$, 1$f_{5/2}$, 2$p_{3/2}$, 0$i_{13/2}$, 1$f_{7/2}$, and~0$h_{9/2}$ between the shell closure $N = 82-126$ and~353 $T = 1$ two-body matrix elements. Calculations are carried out in the hole-hole channel. The~single-particle energies for the valence orbitals (relative to the 2$p_{1/2}$) are as follows: \mbox{$\epsilon(1f_{5/2})$ = 0.570 MeV;} \mbox{$\epsilon(2p_{3/2})$ = 0.898 MeV;} $\epsilon(0i_{13/2})$ = 1.633 MeV; $\epsilon(1f_{7/2})$ = 2.340 MeV; $\epsilon(0h_{9/2})$ = 3.414 MeV~\cite{PhysRevC.94.014312}. For~the selected valence space, there are 21 $T = 1$ monopole matrix elements, for~which strengths are tabulated in Table~\ref{tab:Monopole_Pb}.

\begin{table}[H]
    \caption{{The} strength of the $T = 1$ monopole matrix elements $\bra{j_{\alpha} j_{\beta}} V \ket{j_{\alpha} j_{\beta}}_{J,T}$ of the pb \mbox{effective~interaction.}}
    \label{tab:Monopole_Pb}
    \newcolumntype{C}{>{\centering\arraybackslash}X}
\begin{tabularx}{\textwidth}{CCC}
    \toprule
        \boldmath{$j_{\alpha}$} & \boldmath{$j_{\beta}$} &  \boldmath{$\bra{j_{\alpha} j_{\beta}} V \ket{j_{\alpha} j_{\beta}}_{J,T}$} \\
    \midrule
    2$p_{1/2}$ & 2$p_{1/2}$ & $-$0.0500 \\
    2$p_{1/2}$ & 1$f_{5/2}$ & 0.0504 \\
    2$p_{1/2}$ & 2$p_{3/2}$ & 0.00625 \\
    2$p_{1/2}$ & 0$i_{13/2}$ & 0.0394 \\
    2$p_{1/2}$ & 1$f_{7/2}$ & 0.0467 \\
    2$p_{1/2}$ & 0$h_{9/2}$ & 0.0242 \\
    1$f_{5/2}$ & 1$f_{5/2}$ & 0.00833 \\
    1$f_{5/2}$ & 2$p_{3/2}$ & 0.0241 \\
    1$f_{5/2}$ & 0$i_{13/2}$ & 0.0176 \\
    1$f_{5/2}$ & 1$f_{7/2}$ & 0.0141 \\
    1$f_{5/2}$ & 0$h_{9/2}$ & 0.0886 \\
    2$p_{3/2}$ & 2$p_{3/2}$ & $-$0.0913 \\
    2$p_{3/2}$ & 0$i_{13/2}$ & 0.0822 \\
    2$p_{3/2}$ & 1$f_{7/2}$ & 0.0149 \\
    2$p_{3/2}$ & 0$h_{9/2}$ & 0.0613 \\
    0$i_{13/2}$ & 0$i_{13/2}$ & $-$0.00357 \\
    0$i_{13/2}$ & 1$f_{7/2}$ & 0.114 \\
    0$i_{13/2}$ & 0$h_{9/2}$ & 0.00020 \\
    1$f_{7/2}$ & 1$f_{7/2}$ & $-$0.00661 \\
    1$f_{7/2}$ & 0$h_{9/2}$ & 0.0482 \\
    0$h_{9/2}$ & 0$h_{9/2}$ & 0.0923 \\
    \bottomrule
    \end{tabularx}
\end{table}

In Ref.~\cite{PhysRevC.94.014312}, the~$M$-scheme dimensions for the $M^{\pi} = 0^+$ and the dimensions of the corresponding $J^{\pi} = 0^+$ states in the even-even Pb isotope are shown. In~this work, the~full shell model calculations for $^{194-206}$Pb with the maximum dimension of $3.4 \times 10^9$ were carried~out. 

\subsection{Tin and Xenon~Isotopes}
For the shell model calculations of tin and xenon isotopes, we have considered a model space with the neutron and proton orbitals 0g$_{7/2}$, 1$d_{5/2}$, 1$d_{3/2}$, 2$s_{1/2}$, and~0$h_{11/2}$ lying between the shell closure $Z = N = 50 - 82$, assuming the doubly magic $^{100}$Sn as an inert core. In~this work, we have used the effective interaction derived by applying the Monte Carlo global optimization approach~\cite{PhysRevC.86.044323}. The~effective interaction has been constructed from the realistic CD-Bonn inter-nucleon interaction~\cite{PhysRevC.63.024001} and renormalized by using the perturbative G-matrix approach~\cite{HJORTHJENSEN1995125}. The~core-polarization effects are also taken into account in it. The~mass dependence is not considered in these~calculations. 

\begin{table}[H]
    \setlength\tabcolsep{10.0pt}
    \caption{{The} strength of the $T = 1$ monopole matrix elements $\bra{j_{\alpha} j_{\beta}} V \ket{j_{\alpha} j_{\beta}}_{J,T}$ of the sn100 \mbox{effective~interaction.}}
    \label{tab:Monopole_Sn}
    \newcolumntype{C}{>{\centering\arraybackslash}X}
\begin{tabularx}{\textwidth}{CCC}
    \toprule
    \boldmath{$j_{\alpha}$} & \boldmath{$j_{\beta}$} &  \boldmath{$\bra{j_{\alpha} j_{\beta}} V \ket{j_{\alpha} j_{\beta}}_{J,T}$} \\
    \midrule
    0g$_{7/2}$ & 0g$_{7/2}$ & 0.000 \\
    0g$_{7/2}$ & 1$d_{5/2}$ & $-$0.151 \\
    0g$_{7/2}$ & 1$d_{3/2}$ & $-$0.139 \\
    0g$_{7/2}$ & 2$s_{1/2}$ & $-$0.028 \\
    0g$_{7/2}$ & 0$h_{11/2}$& $-$0.261 \\
    1$d_{5/2}$ & 1$d_{5/2}$ & $-$0.121 \\
    1$d_{5/2}$ & 1$d_{3/2}$ & $-$0.607 \\
    1$d_{5/2}$ & 2$s_{1/2}$ & 0.203 \\
    1$d_{5/2}$ & 0$h_{11/2}$ & $-$0.016 \\
    1$d_{3/2}$ & 1$d_{3/2}$ & 0.179 \\
    1$d_{3/2}$ & 2$s_{1/2}$ & $-$0.768 \\
    1$d_{3/2}$ & 0$h_{11/2}$ & $-$0.116 \\
    2$s_{1/2}$ & 2$s_{1/2}$ & $-$0.749 \\
    2$s_{1/2}$ & 0$h_{11/2}$ & $-$0.013\\
    0$h_{11/2}$& 0$h_{11/2}$ & $-$0.244\\
    \bottomrule
    \end{tabularx}
\end{table}

\textls[-15]{There are 5 single-particle energies and 327 two-body matrix elements for the chosen interactions, 160 of which are $T = 1$ and 167 $T = 0$ elements. There are 15 $T = 1$ monopole terms, which are shown in Table \ref{tab:Monopole_Sn}. The~single-particle energies are the same for both proton and neutron orbitals. The~single-particle energies are taken relative to the 0g$_{7/2}$ orbital as: \mbox{$\epsilon(1d_{5/2})$ = 0.172 MeV;} $\epsilon(1d_{3/2})$ = 5.01279 MeV; $\epsilon(2s_{1/2})$ = 0.36906 MeV; \mbox{$\epsilon(0h_{11/2})$ = 3.24863 MeV~\cite{PhysRevC.86.044323}.}} The~dimensions corresponding to $M^{\pi} = 0^+$ and $J_{\pi} = 0^+$ states in even-even Sn isotopes are shown in Ref.~\cite{PhysRevC.86.044323}. One can notice that the dimension grows drastically with increasing number of valence neutron numbers. The~calculated results with this interaction on Sn isotopes can be found in Ref.~\cite{Sn}.

\section{Results and~Discussion}\label{SecVI}

\subsection{The Energy Spectra of $^{202}$Pb}

\begin{figure}[H]
     \centering
    \includegraphics[width=0.8\linewidth]{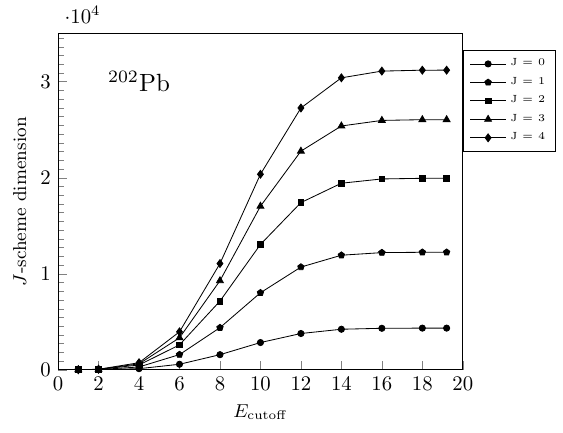}
    \caption{The $J$-scheme dimension for $^{202}$Pb as a function of the specified cutoff energy of the monopole-based truncation for the states with $J = 0$ to $J = 4$.}
    \label{fig:202Pb_dim}
\end{figure}

\begin{figure}[H]
     \centering
    \includegraphics[width=0.85\linewidth]{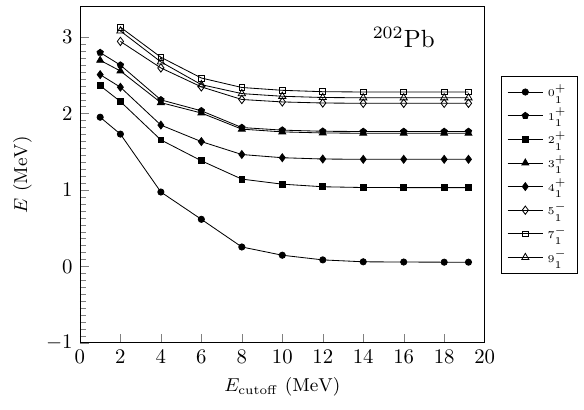}
    \caption{Convergence behaviour of the ground and yrast excited states of $^{202}$Pb as a function of sharp cutoff~energy.}
    \label{fig:Pb_energy_spectra}
\end{figure}

To test the importance of monopole-based and seniority truncations, we study a simple system, $^{202}$Pb, containing six valence holes. The~$M$-scheme dimension for the ground state of $^{202}$Pb is 411,184. In~the first case, we apply monopole truncation with a sharp cutoff criterion in NuShellX~\cite{Nushellx}. In~Figure~\ref{fig:202Pb_dim}, we plot the $J$-scheme dimensions corresponding to spins $J = 0$ to $4$ as a function of energy cutoff. This figure illustrates the exponential growth of the basis states with increasing $E_{\rm cutoff}$. We reach approximately 97\% of the total states at an energy cutoff of 14.0 MeV, as~can be seen from the figure. We have calculated the energies of the low-lying yrast positive- and negative-parity states in the truncated basis states, the~convergence behavior of which is presented in Figure~\ref{fig:Pb_energy_spectra}. At~$E_{\rm cutoff} = 10.0$ MeV, which includes 65\% of the total basis states, we observe that convergence is nearly achieved. The~calculated ground state energy with $E_{\rm cutoff}$ = 10 MeV is 0.147 MeV, which is close to the calculated energy without truncation (0.056 MeV). We found that, with~increasing cutoff energy, the~energies gradually converge to the full-space results. Our results are the same as those obtained from KSHELL~\cite{SHIMIZU2019372}. 

\unskip

\begin{figure}[H]
     \centering
    \includegraphics[width=0.85\linewidth]{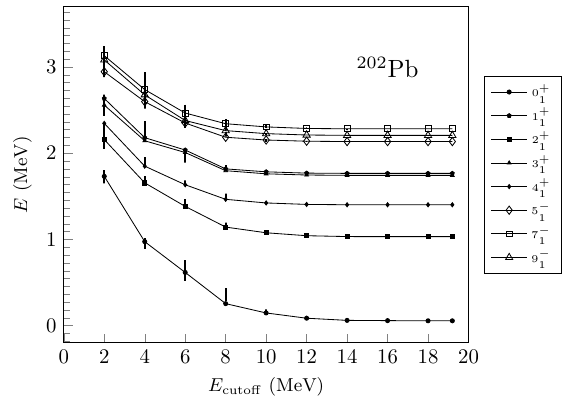}
    \caption{Shell-model energies for the yrast states in $^{202}$Pb with the monopole-based truncation based on sharp cutoff energy and a distribution-type~cutoff.}
    \label{fig:Pb_energy_spectra1}
\end{figure}

Instead of using a sharp cutoff, we implement a distribution-type cutoff (X defined above) in monopole-based truncation for the second case. In~function X, we choose a parameter value of $a = 0.5$. Whether a partition is retained or skipped depends on a random number, so each iteration of the calculation yields a different result. These results are displayed in Figure~\ref{fig:Pb_energy_spectra1} with uncertainties. From~the figure, it is evident that at lower energy cutoffs (below 8 MeV), the~distribution-type cutoff achieves faster convergence compared to the sharp cutoff. For~example, at~$E_{\rm cutoff} = 6$ MeV, the~minimum energy obtained for the $0^+$ state is 0.518 MeV, whereas with the sharp cutoff, the~corresponding energy is 0.617~MeV.

\begin{figure}[H]
\begin{tabular}{cc}
\includegraphics[width=0.5\linewidth]{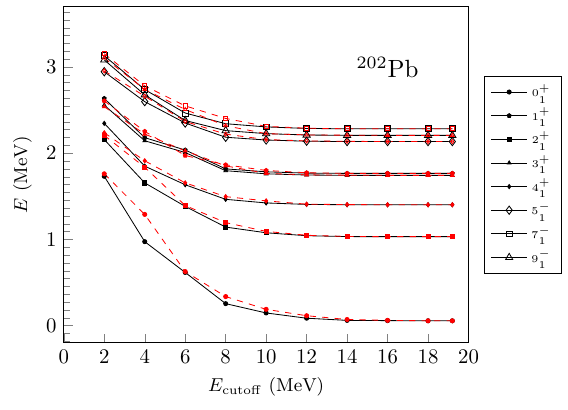}&\includegraphics[width=0.45\linewidth]{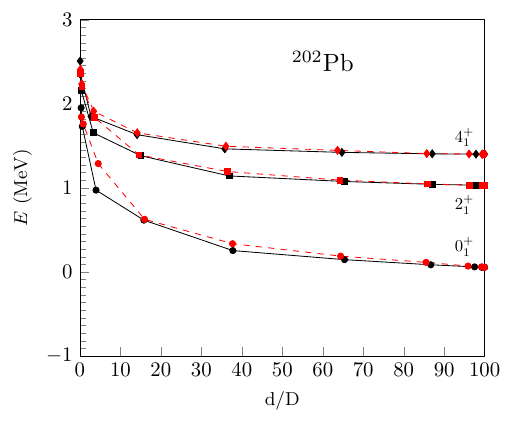}\\
({\bf a})&~~~~~~~~~~~({\bf b})\\
\end{tabular}
	\caption{{(\textbf{a})} 
 Convergence behavior of the ground and yrast excited states of $^{202}$Pb with increasing cutoff energy. (\textbf{b}) Convergence of the shell-model yrast energy states of $^{202}$Pb with respect to the fraction of the $J$-scheme bases~considered.} \label{fig:4}
\end{figure}

As described earlier, the~third type of monopole-based truncation restricts the number of basis states within a given partition according to the function $X$. The~corresponding shell-model results are shown in Figure~\ref{fig:4}a. In~the figure, the~solid lines represent the results obtained using a sharp energy cutoff, while the dashed red lines show the results using this third type of truncation. For~the ground state, the~results with the third type of truncation converge to those from the sharp cutoff at 14.0 MeV. However, for~the excited states, convergence is achieved earlier, at~12 MeV. To~provide a better picture, we have illustrated the convergence of the shell-model energies for the $0_1^+$, $2_1^+$, and~$4_1^+$ states as a function of the fraction of the $J$-scheme basis states considered. In~Figure~\ref{fig:4}b, $D$ represents the total number of $J$-scheme basis states with positive parity, while $d$ denotes the number of basis states included in the truncated shell-model~calculations.

Next, we incorporate the seniority truncation on top of monopole-based truncation for the first time. As~mentioned earlier, we aim to identify the most relevant basis for the ground state by taking the lower seniority states in the $M$-scheme. For~the system considered by us, the~maximum number of unpaired nucleons is six. In~our case, the \mbox{$M$-states} for a partition can have seniority quantum numbers of 0, 2, 4, and~6. We identify the minimum seniority corresponding to a partition and label it as `s.' Then, we define a function that depends on the seniority quantum number such that if the seniority for the random $M$-state within a partition is zero + s (minimum seniority corresponding to that partition), then that particular $M$-state is retained, following which, we proceed for the projection. Seniorities of 2 + s,  4 + s, and~6 + s respectively correspond to 33\%, 66\%, and~99\% chances of generating a new basis state. In~this way, we can reach a physically meaningful starting $M$-basis vector, which is most relevant to the ground~state. 

\begin{figure}[H]
    \begin{subfigure}{0.5\textwidth}
         \includegraphics[width=\textwidth]{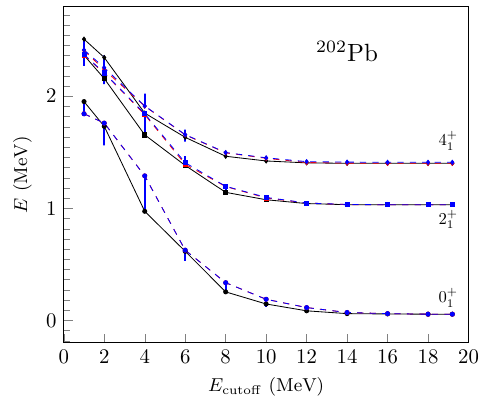}
    \end{subfigure}
     \hfill 
    \begin{subfigure}{0.5\textwidth}
         \includegraphics[width=\textwidth]{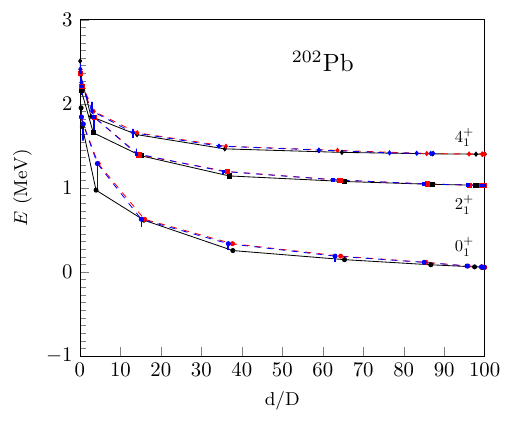}
    \end{subfigure}   
    \caption{{Shell-model} energies for the low-lying states in $^{202}$Pb with monopole-based plus seniority truncations as a function of cutoff energy (\textbf{left}) and a fraction of the bases considered (\textbf{right}).}
    \label{fig:seniority_truncation}
\end{figure}

We have combined the monopole and seniority truncations in this paper for the first time. In~Figure~\ref{fig:seniority_truncation}, we have shown the results of monopole plus seniority truncations for $^{202}$Pb. Compared to the third case, we have further reduced the $J$-basis states based on the seniority quantum number. For~minimum seniority $\leq 2$, we increase the cutoff energy by adding to it 0.5 $\times$ the minimum seniority value. Accordingly, $X$ decreases, which leads to the decrease of the $J$-dimension. The~corresponding results are shown by dashed blue lines in Figure~\ref{fig:seniority_truncation}. 
Then, we implement the seniority truncation in the projection module to start with the $M$-state involving minimum seniority, and the corresponding results are shown in Figure~\ref{fig:seniority_truncation} by uncertainty. It is evident from the figure that seniority truncation improves the results.  
As an illustration, we have displayed the convergence of the energies for the yrast $0^+$, $2^+$, and~$4^+$ states with an increasing number of bases in Figure~\ref{fig:seniority_truncation}. We can clearly see that the results obtained with monopole + seniority truncation are even better at smaller dimensions than the calculation with monopole truncation~alone.

\subsection{The Energy Spectra and $B(E2)$ for $^{106}$Sn}

\begin{figure}[H]
     \centering
    \includegraphics[width=0.85\linewidth]{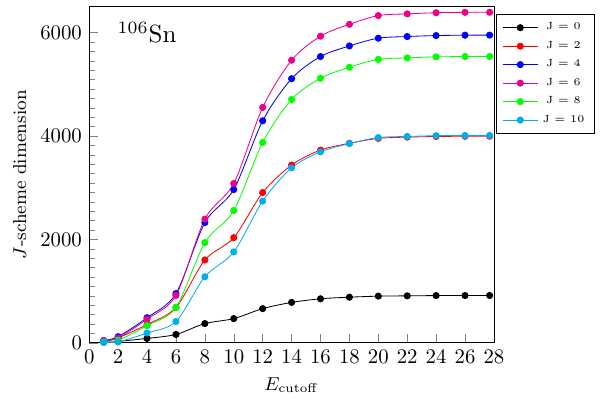}
    \caption{{The} $J$-scheme dimension for $^{106}$Sn as a function of cutoff energy for the states with spins $J = 0-10$.}
    \label{fig:106Sn_dim}
\end{figure}

\begin{figure}[H]
\begin{tabular}{cc}
\hspace{-1em}\includegraphics[width=0.49\linewidth]{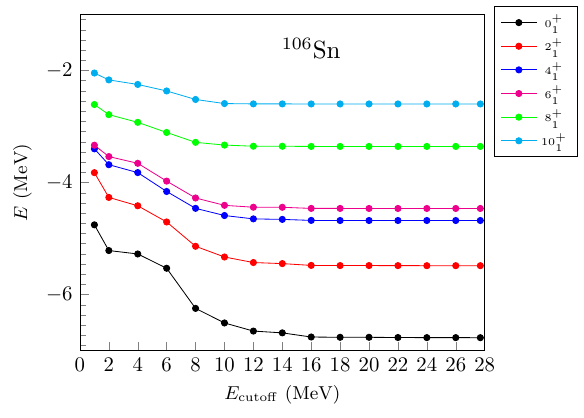}&\includegraphics[width=0.49\linewidth]{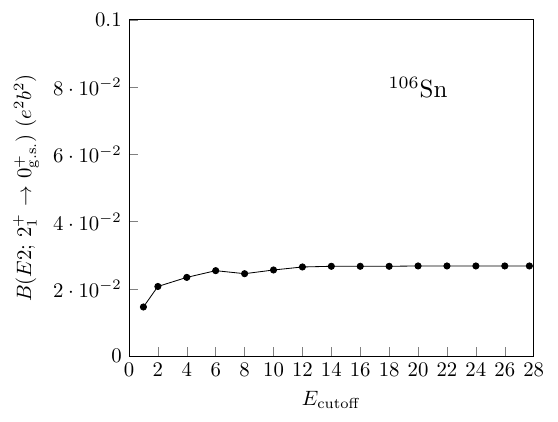}\\
({\bf a})&~~~~~~~~~~~~~~~~~({\bf b})\\
\end{tabular}
\caption{Convergence of (\textbf{a}) the shell model energies for the low-lying  states and (\textbf{b}) the $B(E2)$ strength from the first $2^+$ to $0^+$ in $^{106}$Sn as a function of cutoff~energy.} 
    \label{fig:106Sn}
\end{figure}

We have performed calculations for $^{106}$Sn, 6 valence neutrons on top of the $^{100}$Sn core, using the shell model with the monopole-based truncation. We have shown the $J$-scheme dimension with increasing cutoff energy in Figure~\ref{fig:106Sn_dim}. The~dimension increases rapidly with increasing cutoff energy. When we include all the basis states, we obtain the full-space bases. The~energies of low-lying positive-parity states with even spin are plotted in Figure~\ref{fig:106Sn}a. As~we increase cutoff energy, our truncated results approach the full-space calculation, with~convergence being reached at the cutoff energy of 10 MeV. The~obtained ground state energy at $E_{\rm cutoff} = 10.0$ MeV is \mbox{$-$6.505 MeV,} while the full-space result for the same is \mbox{$-$6.768 MeV.} We noted that a very good convergence is reached with the truncated model space, even though it is only $\approx$ 46\% of the full model~space.

A more stringent test for these truncations would be the calculation of the $E2$ transition strength, since quadrupole degrees of freedom are included through off-diagonal matrix elements (non-monopole term). We have calculated the $B(E2)$ value for the transition between $2_1^+$ and $0^+$ in $^{106}$Sn; corresponding results are shown in Figure~\ref{fig:106Sn}b. The~experimentally measured $B(E2)$ value of the transition $2_1^+$ $\rightarrow$ $0_1^+$ is 0.048 $e^2 b^2$ \cite{PhysRevLett.99.162501}. We have used the neutron effective charge  $e_n^{\rm eff} = 1.0e$ in the $B(E2)$ calculations, the same as used in Ref.~\cite{PhysRevC.72.061305}. The~untruncated calculation gives a $B(E2)$ value of 0.027 $e^2 b^2$. The~discrepancy between theory and experiment might be due to missing contribution from $g_{9/2}$ orbital below $Z = 50$ shell closure~\cite{PhysRevLett.99.162501}. Figure~\ref{fig:106Sn}b shows that we get a converged $B(E2)$ value at small cutoff energy in $^{106}$Sn.

\vspace{-12pt}

\begin{figure}[H]
	\centering
	\includegraphics[width=0.8\linewidth]{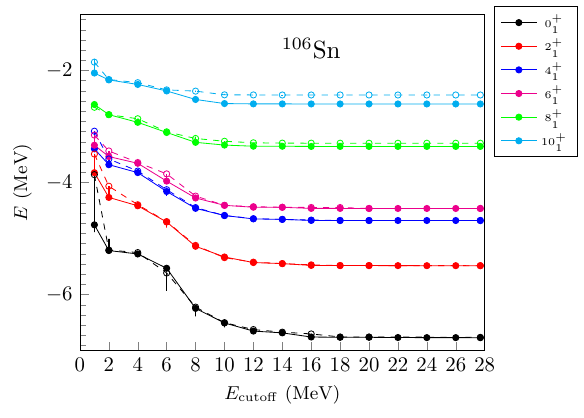}
	\caption{The shell model energies for the low-lying states in $^{106}$Sn with combined monopole and seniority truncations as a function of cutoff~energy.}
	\label{fig:Sn_seniority}
\end{figure}

Similar to the $^{202}$Pb case, we have performed the shell model calculations for low-lying energy states with combined monopole and seniority truncations in $^{106}$Sn. The~results of energies are shown in Figure~\ref{fig:Sn_seniority}. By~imposing the seniority truncation, we can obtain a faster convergence as compared to the monopole-based truncation calculation at small cutoff energy. It is clear from the figure that for the high-spin states, states with high seniority are required to get the full-space~results.

\subsection{The Energy Spectra and $B(E2)$ for $^{108}$Xe}
In the previous sections, we performed calculations for a system containing only one type of nucleon. Now, we aim to test the effectiveness of monopole-based truncation in a more complex system involving both protons and neutrons. For~this purpose, we have done truncated calculations for the nucleus $^{108}$Xe, with~equal numbers of protons and neutrons ($N = Z = 54$). This nucleus is predicted to lie on the proton-drip line. Experimentally, only the ground state has been measured, which corresponds to a $0^+$ state~\cite{MOLLER20191, NNDC}.

\begin{figure}[H]
\begin{tabular}{cc}
\hspace{-1em}\includegraphics[width=0.48\linewidth]{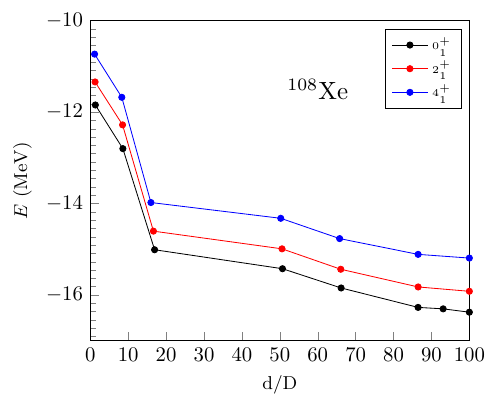}&\includegraphics[width=0.51\linewidth]{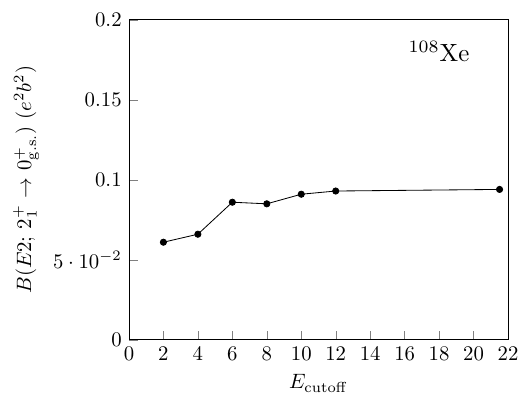}\\
({\bf a})&~~~~~~~~~~~~~~~~~~~~({\bf b})\\
\end{tabular}
\caption{(\textbf{a}) Energies of the lowest three states of $^{108}$Xe as a function of the fraction of the bases taken into consideration. (\textbf{b}) Calculated $B(E2)$ value for the transition $2_1^+$ $\rightarrow$ $0^+_{\rm g.s.}$ in $^{108}$Xe against the cutoff energy of the monopole~truncation.}
\label{fig:108xe}
\end{figure}

In our calculations, we have used the same interaction (sn100) as employed in the studies of Sn isotopes. The $M$-scheme dimension for the ground state of $^{108}$Xe is \mbox{7.4 $\times$ $10^7$.} In addition to the ground state, we provide predictions for the first excited $2^+$ and $4^+$ states, which could be useful for future experimental measurements. \mbox{Figure \ref{fig:108xe}a} shows the calculated energies of the three lowest-lying positive parity states ($0^+$, $2^+$, and $4^+$) in $^{108}$Xe. Without truncation, the exact energies for these states are found to be $-$16.377 MeV, $-$15.920 MeV, and $-$15.193 MeV, respectively. We then applied monopole-based truncation using a sharp cutoff criterion. With an energy cutoff of $E_{\rm cutoff} = 6.0$ MeV, the calculated energies are $-$15.011 MeV, $-$14.604 MeV, and $-$13.978 MeV for the $0^+$, $2^+$, and $4^+$ states, respectively. Calculations are then performed with $E_{\rm cutoff}= 8.0$ MeV, which corresponds to approximately 50\% of the total model space. In this case, the energies were $-$15.426 MeV, $-$14.991 MeV, and $-$14.323 MeV for the above-mentioned states. Remarkably, convergence is achieved using less than 20\% of the full $J$-scheme basis, demonstrating the efficiency of the monopole-based truncation method even in systems with both types of nucleons. In Figure \ref{fig:108xe}b, the convergence behavior of $B(E2)$ for the transition $2_1^+$ $\rightarrow$ $0^+_{\rm g.s.}$ is presented with increasing cutoff energy. There is no experimental $B(E2)$ value available yet. The $B(E2)$ strength for the full-space calculation is obtained as 940.6 $e^2 {\rm fm}^4$. From the figure, it appears that convergence for $E2$ strength can be reached at a small part of the full-space basis. This indicates that the monopole-based truncation gives a well-converged value of the observables with a reduced dimension of the calculation. 

In NuShellX, we have implemented the cutoff criteria separately for protons and neutrons, since projection is done separately for them. On~the other hand, in~KSHELL, the~monopole-based truncation is made on the total monopole Hamiltonian, which contains the proton-neutron part as well. The~monopole-based truncation with \mbox{$E_{\rm cutoff} = 8.0$ MeV} yields the energies of $-$14.795, $-$14.376, and $-$13.761 MeV for the $0^+$, $2^+$, and~$4^+$ states, respectively. It indicates that, for~the system with both kinds of valence nucleons, calculated results are different from these two shell-model codes. Recently, shell model calculations have been performed for $^{200}$Hg without truncation and $^{199,199}\mathrm{Hg}$ with monopole-based truncation at $E_{\rm cutoff} = 12$ MeV and other Hg isotopes at $E_{\rm cutoff} = 10$ MeV using the KSHELL~\cite{PhysRevC.110.024306}. 

\subsection{The Energy Spectra and $B(E2)$ for $^{110}$Xe}
We now extend our study to a more complex system, the~extremely neutron-deficient $N = Z + 2$ nucleus, $^{110}$Xe, which contains an unequal number of protons and neutrons \linebreak (4 valence protons and 6 valence neutrons). The~$M$-scheme dimension for the ground state of this nucleus is 1.7 $\times$ $10^9$, which is the maximum dimension that current shell-model codes can handle. In~Ref.~\cite{PhysRevLett.99.022501}, the~energies of the three lowest states in $^{110}$Xe have been experimentally observed. The~excitation energies of the tentative $2^+$ and $4^+$ states were measured to be 470 keV and 1113 keV, respectively. Interestingly, the~results from Ref.~\cite{PhysRevLett.99.022501} revealed a deviation from the normal trend of increasing energies with decreasing neutron number, which was attributed to enhanced collectivity. This anomaly is suggested to arise from the $T = 0$ neutron-proton interaction in nuclei near the $N = Z$ region.

\begin{figure}[H]
    \begin{subfigure}{0.5\textwidth}
    \includegraphics[width=\textwidth]{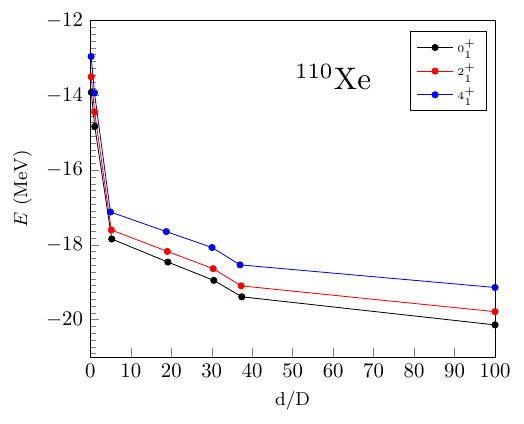}        
    \end{subfigure}
    \hfil
    \begin{subfigure}{0.5\textwidth}
    \includegraphics[width=\textwidth]{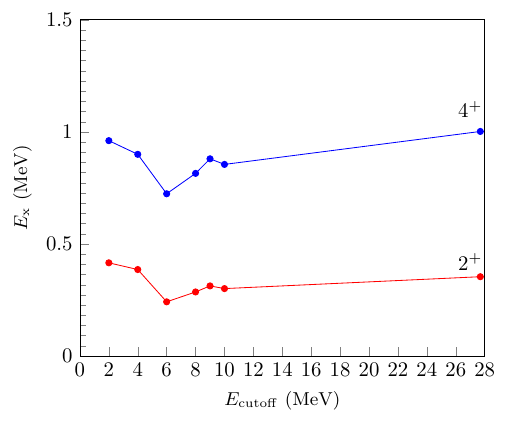}        
    \end{subfigure}
    \caption{{(\textbf{Left})}
 Energies of the lowest three states of $^{110}$Xe as a function of the fraction of the bases taken into consideration. (\textbf{Right}) Excitation energies of first $2^+$ and $4^+$ states relative to the ground state as a function of $E_{\rm cutoff}$.}
    \label{fig:110xe}
\end{figure}

\begin{figure}[H]
	\centering
	\includegraphics[width=0.6\textwidth]{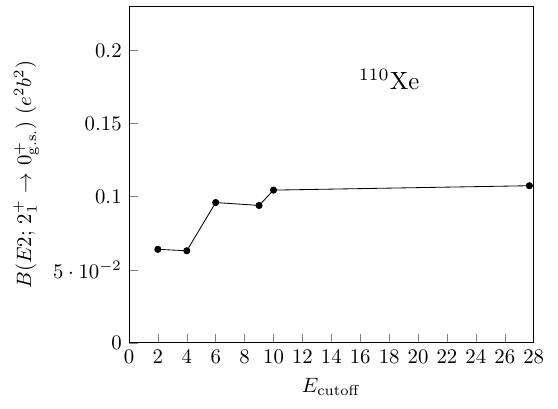}
	\caption{Convergence of the $B(E2)$ value for $^{110}$Xe with the cutoff energy of the monopole-based~truncation.}
	\label{fig:110Xe_BE2}
\end{figure}

\textls[-5]{We have performed full-space shell-model calculations using the sn100 interaction for $^{110}$Xe, and the obtained energies for the lowest three states are $-$20.143 MeV,  \mbox{$-$19.789 MeV,} and  $-$19.141 MeV. The calculated excitation energies for the $2^+$ and $4^+$ states are \mbox{354 keV} and 1002 keV, respectively, which show good agreement with the experimental \mbox{data \cite{PhysRevLett.99.022501}.}} The calculations considering monopole-based truncation have also been performed. \mbox{Figure \ref{fig:110xe}} shows the convergence of the three low-lying states for $^{110}$Xe as a function of the fraction of the $J$-scheme bases taken into account. At a cutoff energy of 2 MeV, the obtained ground state energy is  $-$13.919 MeV. Hence, the difference between the full space calculation and the calculation with $E_{\rm cutoff}$ = 2 MeV is 6.224 MeV. For the calculation considering $\sim$37\% of the total dimension ($E_{\rm cutoff}$ = 10 MeV), the difference is reduced to 752 keV. With increasing cutoff energy, the energy of the ground state approaches the exact value. NuShellX gives the results of $-$19.391,  $-$19.098, and $-$18.536 MeV for $0^+$, $2^+$, and $4^+$ states, respectively, while the corresponding energies are  $-$18.750, $-$18.436, and $-$17.887 MeV with KSHELL at $E_{\rm cutoff} = 10.0$ MeV. The convergence is even faster when focusing solely on the spectrum relative to the ground state, as shown in Figure \ref{fig:110xe}. There is a one-to-one correspondence between d/D and the cutoff energy in the left and right panels of Figure \ref{fig:110xe}. The convergence is already achieved by taking only around 37\% of the total bases. These calculations are significantly faster than the full space calculation. We have checked the predictive power of the monopole-based truncation by  calculating the reduced $E2$ transition strength for the open-shell nucleus $^{110}$Xe (with 4 valence protons and 6 valence neutrons). In \mbox{Figure \ref{fig:110Xe_BE2},} we have plotted the $B(E2)$ between first $2^+$ and $0^+$ states for $^{110}$Xe. Although it is expected that achieving $B(E2)$ convergence is challenging for deformed nuclei, our results show a well-converged value is achieved at 10 MeV cutoff energy. This indicates that the monopole-based truncation with small cutoff energy is sufficient in reproducing the full-space transition strength for the open-shell nuclei.

\vspace{-12pt}

The monopole and seniority truncations are straightforward to implement in the shell model code. Based on the results of low-lying energy spectra and reduced $E2$ transition strength, we can conclude that these truncations are among the most efficient truncations (without breaking symmetries) for~reducing the computational demands of large-scale \mbox{shell-model} calculations. Our results show that considering $\sim$10--30\% of the total dimension gives a reasonable result for both spherical and deformed nuclei. These truncations would be beneficial for the systems with many valence
nucleons, allowing for feasible and accurate calculations of the nuclear structure~properties. 

\section{Conclusions}\label{SecVII}
In the present work, we have introduced two truncation techniques for the configuration interaction shell model approach to tackle the huge dimension of the Hamiltonian matrix: monopole-based truncation and seniority truncation inspired by strong pairing correlation. These truncations incorporate the selection of certain partitions according to their importance and retain $M$-basis states with lower seniority within each partition. As~a benchmark, we have carried out large-scale shell model calculations for various systems, including $^{106}$Sn, $^{108,110}$Xe, and~$^{202}$Pb, comparing the calculated energies of ground and excited states and $B(E2)$ values in both truncated and full basis spaces. A~good convergence is obtained with a small part of the total bases, which demonstrates the effectiveness of these truncations in reducing the dimension. These truncation methods hold promise for extending the reach of shell-model calculations to heavier nuclei, where the conventional shell model becomes computationally~challenging.


\vspace{6pt}

\authorcontributions{{P.C.: Performed calculations, investigation, writing---review and editing.\\
		  C.Q.: Conceptualization, methodology, investigation, writing---review and editing.} 
}

\funding{{{The research was funded by Olle Engkvists Foundation.}}
}

\acknowledgments{
  The~computations were enabled by resources at PDC Center for High Performance Computing, KTH Royal Institute of Technology provided by the National Academic Infrastructure for Supercomputing in Sweden (NAISS). }

\begin{adjustwidth}{-\extralength}{0cm}

\reftitle{References}

\end{adjustwidth}

\begin{thebibliography}{999}

\bibitem[Hergert et~al.(2016)Hergert, Bogner, Morris, Schwenk, and
  Tsukiyama]{HERGERT2016165}
Hergert, H.; Bogner, S.; Morris, T.; Schwenk, A.; Tsukiyama, K.
\newblock The In-Medium Similarity Renormalization Group: A novel ab~initio
  method for nuclei.
\newblock {\em Phys. Rep.} {\bf 2016}, {\em 621},~165--222.
\newblock
  {\url{https://doi.org/10.1016/j.physrep.2015.12.007}}.

\bibitem[Stroberg et~al.(2016)Stroberg, Hergert, Holt, Bogner, and
  Schwenk]{PhysRevC.93.051301}
Stroberg, S.R.; Hergert, H.; Holt, J.D.; Bogner, S.K.; Schwenk, A.
\newblock Ground and excited states of doubly open-shell nuclei from ab~initio
  valence-space Hamiltonians.
\newblock {\em Phys. Rev. C} {\bf 2016}, {\em 93},~051301.
\newblock {\url{https://doi.org/10.1103/PhysRevC.93.051301}}.

\bibitem[Parzuchowski et~al.(2017)Parzuchowski, Stroberg, Navr\'atil, Hergert,
  and Bogner]{PhysRevC.96.034324}
Parzuchowski, N.M.; Stroberg, S.R.; Navr\'atil, P.; Hergert, H.; Bogner, S.K.
\newblock Ab~initio electromagnetic observables with the in-medium similarity
  renormalization group.
\newblock {\em Phys. Rev. C} {\bf 2017}, {\em 96},~034324.
\newblock {\url{https://doi.org/10.1103/PhysRevC.96.034324}}.

\bibitem[Stroberg et~al.(2022)Stroberg, Henderson, Hackman, Ruotsalainen,
  Hagen, and Holt]{PhysRevC.105.034333}
Stroberg, S.R.; Henderson, J.; Hackman, G.; Ruotsalainen, P.; Hagen, G.; Holt,
  J.D.
\newblock Systematics of $E2$ strength in the $sd$ shell with the valence-space
  in-medium similarity renormalization group.
\newblock {\em Phys. Rev. C} {\bf 2022}, {\em 105},~034333.
\newblock {\url{https://doi.org/10.1103/PhysRevC.105.034333}}.

\bibitem[Pieper et~al.(2001)Pieper, Pandharipande, Wiringa, and
  Carlson]{PhysRevC.64.014001}
Pieper, S.C.; Pandharipande, V.R.; Wiringa, R.B.; Carlson, J.
\newblock Realistic models of pion-exchange three-nucleon interactions.
\newblock {\em Phys. Rev. C} {\bf 2001}, {\em 64},~014001.
\newblock {\url{https://doi.org/10.1103/PhysRevC.64.014001}}.

\bibitem[Pervin et~al.(2007)Pervin, Pieper, and Wiringa]{PhysRevC.76.064319}
Pervin, M.; Pieper, S.C.; Wiringa, R.B.
\newblock Quantum Monte Carlo calculations of electroweak transition matrix
  elements in $A=6,7$ nuclei.
\newblock {\em Phys. Rev. C} {\bf 2007}, {\em 76},~064319.
\newblock {\url{https://doi.org/10.1103/PhysRevC.76.064319}}.

\bibitem[Marcucci et~al.(2008)Marcucci, Pervin, Pieper, Schiavilla, and
  Wiringa]{PhysRevC.78.065501}
Marcucci, L.E.; Pervin, M.; Pieper, S.C.; Schiavilla, R.; Wiringa, R.B.
\newblock Quantum Monte Carlo calculations of magnetic moments and $M1$
  transitions in $A\ensuremath{\leqslant}7$ nuclei including meson-exchange
  currents.
\newblock {\em Phys. Rev. C} {\bf 2008}, {\em 78},~065501.
\newblock {\url{https://doi.org/10.1103/PhysRevC.78.065501}}.

\bibitem[Jansen et~al.(2014)Jansen, Engel, Hagen, Navratil, and
  Signoracci]{PhysRevLett.113.142502}
Jansen, G.R.; Engel, J.; Hagen, G.; Navratil, P.; Signoracci, A.
\newblock Ab~Initio Coupled-Cluster Effective Interactions for the Shell Model:
  Application to Neutron-Rich Oxygen and Carbon Isotopes.
\newblock {\em Phys. Rev. Lett.} {\bf 2014}, {\em 113},~142502.
\newblock {\url{https://doi.org/10.1103/PhysRevLett.113.142502}}.

\bibitem[Hagen et~al.(2010)Hagen, Papenbrock, Dean, and
  Hjorth-Jensen]{PhysRevC.82.034330}
Hagen, G.; Papenbrock, T.; Dean, D.J.; Hjorth-Jensen, M.
\newblock Ab~initio coupled-cluster approach to nuclear structure with modern
  nucleon-nucleon interactions.
\newblock {\em Phys. Rev. C} {\bf 2010}, {\em 82},~034330.
\newblock {\url{https://doi.org/10.1103/PhysRevC.82.034330}}.

\bibitem[Hagen et~al.(2016)Hagen, Hjorth-Jensen, Jansen, and
  Papenbrock]{Hagen2016}
Hagen, G.; Hjorth-Jensen, M.; Jansen, G.R.; Papenbrock, T.
\newblock Emergent properties of nuclei from ab~initio coupled-cluster
  calculations.
\newblock {\em Phys. Scr.} {\bf 2016}, {\em 91},~063006.
\newblock {\url{https://doi.org/10.1088/0031-8949/91/6/063006}}.

\bibitem[Barrett et~al.(2013)Barrett, Navrátil, and Vary]{BARRETT2013131}
Barrett, B.R.; Navrátil, P.; Vary, J.P.
\newblock Ab~initio no core shell model.
\newblock {\em Prog. Part. Nucl. Phys.} {\bf 2013}, {\em
  69},~131--181.
\newblock {\url{https://doi.org/10.1016/j.ppnp.2012.10.003}}.

\bibitem[Epelbaum et~al.(2011)Epelbaum, Krebs, Lee, and
  Mei\ss{}ner]{PhysRevLett.106.192501}
Epelbaum, E.; Krebs, H.; Lee, D.; Mei\ss{}ner, U.G.
\newblock Ab~Initio Calculation of the Hoyle State.
\newblock {\em Phys. Rev. Lett.} {\bf 2011}, {\em 106},~192501.
\newblock {\url{https://doi.org/10.1103/PhysRevLett.106.192501}}.

\bibitem[SHAVITT(1998)]{doi:10.1080/002689798168303}
SHAVITT, I.
\newblock The history and evolution of configuration interaction.
\newblock {\em Mol. Phys.} {\bf 1998}, {\em 94},~3--17.
\newblock {\url{https://doi.org/10.1080/002689798168303}}.

\bibitem[Brussaard and Glaudemans(1977)]{brussaard1977shell}
Brussaard, P.J.; Glaudemans, P.W.M.
\newblock {\em Shell-Model Applications in Nuclear Spectroscopy};
\newblock North-Holland Publishing Company: Amsterdam, The Netherlands, 1977.

\bibitem[Caurier et~al.(2005)Caurier, Mart\'{\i}nez-Pinedo, Nowacki, Poves, and
  Zuker]{Caurier2005427}
Caurier, E.; Mart\'{\i}nez-Pinedo, G.; Nowacki, F.; Poves, A.; Zuker, A.P.
\newblock The shell model as a unified view of nuclear structure.
\newblock {\em Rev. Mod. Phys.} {\bf 2005}, {\em 77},~427--488.
\newblock {\url{https://doi.org/10.1103/RevModPhys.77.427}}.

\bibitem[Brown and
  Wildenthal(1988)]{annurev:/content/journals/10.1146/annurev.ns.38.120188.000333}
Brown, B.A.; Wildenthal, B.H.
\newblock Status of the Nuclear Shell Model.
\newblock {\em Annu. Rev. Nucl. Part. Sci.} {\bf 1988}, {\em
  38},~29--66.
\newblock
  {\url{https://doi.org/10.1146/annurev.ns.38.120188.000333}}.

\bibitem[Tsunoda and Otsuka(2020)]{Tsunoda2020}
Tsunoda, Y.; Otsuka, T., Configuration Interaction Approach to Atomic Nuclei:
  The Shell Model.
\newblock In {\em Handbook of Nuclear Physics}; Tanihata, I., Toki, H., Kajino,
  T., Eds.; Springer Nature: Singapore, 2020; pp. 1--49.
\newblock {\url{https://doi.org/10.1007/978-981-15-8818-1_17-1}}.

\bibitem[Heyde(1998)]{heyde1998nucleons}
Heyde, K.
\newblock {\em From Nucleons to the Atomic Nucleus: Perspectives in Nuclear
  Physics}; Springer: {Berlin/Heidelberg, Germany}
,  1998.

\bibitem[Roth and Langhammer(2010)]{ROTH2010272}
Roth, R.; Langhammer, J.
\newblock Padé-resummed high-order perturbation theory for nuclear structure
  calculations.
\newblock {\em Phys. Lett. B} {\bf 2010}, {\em 683},~272--277.
\newblock
  {\url{https://doi.org/10.1016/j.physletb.2009.12.046}}.

\bibitem[Tichai et~al.(2018)Tichai, Arthuis, Duguet, Hergert, Somà, and
  Roth]{TICHAI2018195}
Tichai, A.; Arthuis, P.; Duguet, T.; Hergert, H.; Somà, V.; Roth, R.
\newblock Bogoliubov many-body perturbation theory for open-shell nuclei.
\newblock {\em Phys. Lett. B} {\bf 2018}, {\em 786},~195--200.
\newblock
  {\url{https://doi.org/10.1016/j.physletb.2018.09.044}}.

\bibitem[Honma et~al.(1995)Honma, Mizusaki, and Otsuka]{PhysRevLett.75.1284}
Honma, M.; Mizusaki, T.; Otsuka, T.
\newblock Diagonalization of Hamiltonians for Many-Body Systems by Auxiliary
  Field Quantum Monte Carlo Technique.
\newblock {\em Phys. Rev. Lett.} {\bf 1995}, {\em 75},~1284--1287.
\newblock {\url{https://doi.org/10.1103/PhysRevLett.75.1284}}.

\bibitem[Otsuka et~al.(1998)Otsuka, Honma, and Mizusaki]{PhysRevLett.81.1588}
Otsuka, T.; Honma, M.; Mizusaki, T.
\newblock Structure of the
  $\mathit{N}\mathit{}=\mathit{}\mathit{Z}\phantom{\rule{0ex}{0ex}}=\phantom{\rule{0ex}{0ex}}28$
  Closed Shell Studied by Monte Carlo Shell Model Calculation.
\newblock {\em Phys. Rev. Lett.} {\bf 1998}, {\em 81},~1588--1591.
\newblock {\url{https://doi.org/10.1103/PhysRevLett.81.1588}}.

\bibitem[Shimizu(2022)]{physics4030071}
Shimizu, N.
\newblock Recent Progress of Shell-Model Calculations, Monte Carlo Shell Model,
  and Quasi-Particle Vacua Shell Model.
\newblock {\em Physics} {\bf 2022}, {\em 4},~1081--1093.
\newblock {\url{https://doi.org/10.3390/physics4030071}}.

\bibitem[Shimizu and Mizusaki(2018)]{PhysRevC.98.054309}
Shimizu, N.; Mizusaki, T.
\newblock Variational Monte Carlo method for shell-model calculations in
  odd-mass nuclei and restoration of symmetry.
\newblock {\em Phys. Rev. C} {\bf 2018}, {\em 98},~054309.
\newblock {\url{https://doi.org/10.1103/PhysRevC.98.054309}}.

\bibitem[Gao(2022)]{GAO2022136795}
Gao, Z.C.
\newblock Variation after projection calculations for high-spin states.
\newblock {\em Phys. Lett. B} {\bf 2022}, {\em 824},~136795.
\newblock
  {\url{https://doi.org/10.1016/j.physletb.2021.136795}}.

\bibitem[Lin et~al.(2024)Lin, Zhou, Yao, and Hergert]{sym16040409}
Lin, W.; Zhou, E.; Yao, J.; Hergert, H.
\newblock Quantum-Number Projected Generator Coordinate Method for 21Ne with a
  Chiral Two-Nucleon-Plus-Three-Nucleon Interaction.
\newblock {\em Symmetry} {\bf 2024}, {\em 16}, {409} 

\bibitem[Mayer(1948)]{PhysRev.74.235}
Mayer, M.G.
\newblock On Closed Shells in Nuclei.
\newblock {\em Phys. Rev.} {\bf 1948}, {\em 74},~235--239.
\newblock {\url{https://doi.org/10.1103/PhysRev.74.235}}.

\bibitem[Mayer(1949)]{PhysRev.75.1969}
Mayer, M.G.
\newblock On Closed Shells in Nuclei. II.
\newblock {\em Phys. Rev.} {\bf 1949}, {\em 75},~1969--1970.
\newblock {\url{https://doi.org/10.1103/PhysRev.75.1969}}.

\bibitem[Haxel et~al.(1949)Haxel, Jensen, and Suess]{PhysRev.75.1766.2}
Haxel, O.; Jensen, J.H.D.; Suess, H.E.
\newblock On the ``Magic Numbers'' in Nuclear Structure.
\newblock {\em Phys. Rev.} {\bf 1949}, {\em 75},~1766--1766.
\newblock {\url{https://doi.org/10.1103/PhysRev.75.1766.2}}.

\bibitem[Mayer(1950)]{PhysRev.78.16}
Mayer, M.G.
\newblock Nuclear Configurations in the Spin-Orbit Coupling Model. I. Empirical
  Evidence.
\newblock {\em Phys. Rev.} {\bf 1950}, {\em 78},~16--21.
\newblock {\url{https://doi.org/10.1103/PhysRev.78.16}}.

\bibitem[Kuo and Brown(1966)]{KUO196640}
Kuo, T.; Brown, G.
\newblock Structure of finite nuclei and the free nucleon-nucleon interaction:
  An application to 18O and 18F.
\newblock {\em Nucl. Phys.} {\bf 1966}, {\em 85},~40--86.
\newblock {\url{https://doi.org/10.1016/0029-5582(66)90131-3}}.

\bibitem[Honma et~al.(2002)Honma, Otsuka, Brown, and
  Mizusaki]{PhysRevC.65.061301}
Honma, M.; Otsuka, T.; Brown, B.A.; Mizusaki, T.
\newblock Effective interaction for $\mathrm{pf}$-shell nuclei.
\newblock {\em Phys. Rev. C} {\bf 2002}, {\em 65},~061301.
\newblock {\url{https://doi.org/10.1103/PhysRevC.65.061301}}.

\bibitem[Cohen and Kurath(1965)]{COHEN19651}
Cohen, S.; Kurath, D.
\newblock Effective interactions for the 1p shell.
\newblock {\em Nucl. Phys.} {\bf 1965}, {\em 73},~1--24.
\newblock {\url{https://doi.org/10.1016/0029-5582(65)90148-3}}.

\bibitem[Brown and Richter(2006)]{PhysRevC.74.034315}
Brown, B.A.; Richter, W.A.
\newblock New ``USD'' Hamiltonians for the ${\mathit{sd}}$ shell.
\newblock {\em Phys. Rev. C} {\bf 2006}, {\em 74},~034315.
\newblock {\url{https://doi.org/10.1103/PhysRevC.74.034315}}.

\bibitem[Yuan et~al.(2012)Yuan, Suzuki, Otsuka, Xu, and
  Tsunoda]{PhysRevC.85.064324}
Yuan, C.; Suzuki, T.; Otsuka, T.; Xu, F.; Tsunoda, N.
\newblock Shell-model study of boron, carbon, nitrogen, and oxygen isotopes
  with a monopole-based universal interaction.
\newblock {\em Phys. Rev. C} {\bf 2012}, {\em 85},~064324.
\newblock {\url{https://doi.org/10.1103/PhysRevC.85.064324}}.

\bibitem[Cole et~al.(1975)Cole, Watt, and Whitehead]{COLE197524}
Cole, B.; Watt, A.; Whitehead, R.
\newblock Three-body forces and shell-model binding energies.
\newblock {\em Phys. Lett. B} {\bf 1975}, {\em 57},~24--26.
\newblock {\url{https://doi.org/10.1016/0370-2693(75)90234-8}}.

\bibitem[Andreozzi and Sartoris(1976)]{ANDREOZZI1976388}
Andreozzi, F.; Sartoris, G.
\newblock Effective three-body interactions in the f72 shell.
\newblock {\em Nucl. Phys. A} {\bf 1976}, {\em 270},~388--398.
\newblock {\url{https://doi.org/10.1016/0375-9474(76)90452-8}}.

\bibitem[Lanczos(1950)]{Lanczos:1950zz}
Lanczos, C.
\newblock {An iteration method for the solution of the eigenvalue problem of
  linear differential and integral operators}.
\newblock {\em J. Res. Natl. Bur. Stand. Sect. B} {\bf 1950}, {\em 45},~255--282.
\newblock {\url{https://doi.org/10.6028/jres.045.026}}.

\bibitem[Johnson et~al.(2013)Johnson, Ormand, and Krastev]{JOHNSON20132761}
Johnson, C.W.; Ormand, W.E.; Krastev, P.G.
\newblock Factorization in large-scale many-body calculations.
\newblock {\em Comput. Phys. Commun.} {\bf 2013}, {\em
  184},~2761--2774.
\newblock {\url{https://doi.org/10.1016/j.cpc.2013.07.022}}.

\bibitem[Shimizu et~al.(2019)Shimizu, Mizusaki, Utsuno, and
  Tsunoda]{SHIMIZU2019372}
Shimizu, N.; Mizusaki, T.; Utsuno, Y.; Tsunoda, Y.
\newblock Thick-restart block Lanczos method for large-scale shell-model
  calculations.
\newblock {\em Comput. Phys. Commun.} {\bf 2019}, {\em
  244},~372--384.
\newblock {\url{https://doi.org/10.1016/j.cpc.2019.06.011}}.

\bibitem[deShalit and Talmi(1963)]{book:91951242}
deShalit, A.; Talmi, I.
\newblock {\em Nuclear Shell Theory}; Dover: {Mineola, NY, USA},  1963.

\bibitem[Xu et~al.(2012)Xu, Qi, Blomqvist, Liotta, and Wyss]{XU201251}
Xu, Z.; Qi, C.; Blomqvist, J.; Liotta, R.; Wyss, R.
\newblock Multistep shell model description of spin-aligned neutron--proton
  pair coupling.
\newblock {\em Nucl. Phys. A} {\bf 2012}, {\em 877},~51--58.
\newblock
  {\url{https://doi.org/10.1016/j.nuclphysa.2011.12.005}}.

\bibitem[Jiao et~al.(2014)Jiao, Sun, Xu, Xu, and Qi]{PhysRevC.90.024306}
Jiao, L.F.; Sun, Z.H.; Xu, Z.X.; Xu, F.R.; Qi, C.
\newblock Correlated-basis method for shell-model calculations.
\newblock {\em Phys. Rev. C} {\bf 2014}, {\em 90},~024306.
\newblock {\url{https://doi.org/10.1103/PhysRevC.90.024306}}.

\bibitem[L\"OWDIN(1964)]{RevModPhys.36.966}
L\"OWDIN, P.O.
\newblock Angular Momentum Wavefunctions Constructed by Projector Operators.
\newblock {\em Rev. Mod. Phys.} {\bf 1964}, {\em 36},~966--976.
\newblock {\url{https://doi.org/10.1103/RevModPhys.36.966}}.

\bibitem[Langanke et~al.(1991)Langanke, Maruhn, and Koonin]{Langanke:1991gox}
Langanke, K.; Maruhn, J.A.; Koonin, S.E. (Eds.)
\newblock {\em {Computational Nuclear Physics 1: Nuclear Structure}}; Springer: {Berlin/Heidelberg, Germany} 
, 1991.
\newblock {\url{https://doi.org/10.1007/978-3-642-76356-4}}.

\bibitem[Qi and Xu()]{qi2007modernshellmodeldiagonalizationsrealistic}
Qi, C.; Xu, F.R. {Modern shell-model diagonalizations with realistic NN forces.} 
 \emph{{arXiv}} \textbf{2007}, arXiv:nucl-th/0701036. \url{https://doi.org/10.48550/arXiv.nucl-th/0701036}. 

\bibitem[{Caurier} and {Nowacki}(1999)]{etde_359035}
{Caurier}, E.; {Nowacki}, F.
\newblock {Present Status of Shell Model Techniques}.
\newblock {\em Acta Phys. Pol. B} {\bf 1999}, {\em 30},~705.

\bibitem[Rae()]{Nushellx}
Rae, W.D.M.
\newblock {NuShellX}.  Available online: \url{http://www.garsington.eclipse.co.uk/} ({20 Oct. 2023}) 
{; there is a well
	maintained {MSU} version, as well as a {KTH} version available on request.} 



\bibitem{PhysRevC.50.R2274}
Horoi, M.; Brown, B.A.; Zelevinsky, V.
\newblock {Truncation method for shell model calculations.}
\newblock {\em Phys. Rev. C} {\bf 1994}, {\em 50},~R2274--R2277
\newblock {\url{https://doi.org/10.1103/PhysRevC.50.R2274}}

\bibitem{PhysRevLett.82.2064}
Horoi, M.; Volya, A.; Zelevinsky, V.
\newblock {Chaotic Wave Functions and Exponential Convergence of Low-Lying Energy Eigenvalues}
\newblock {\em Phys. Rev. Lett.} {\bf 1999}, {\em 82},~2064
\newblock {\url{https://doi.org/10.1103/PhysRevLett.82.2064}}

\bibitem{PhysRevC.65.027303}
Horoi, M.; Brown, B.A.; Zelevinsky, V.
\newblock {Applying the exponential convergence method: Shell-model binding energies of ${0f}_{7/2}$ nuclei relative to ${}^{40}\mathrm{Ca}$}
\newblock {\em Phys. Rev. C} {\bf 2002}, {\em 65},~027303
\newblock {\url{https://doi.org/10.1103/PhysRevC.65.027303}}

\bibitem[Dukelsky et~al.(2002)Dukelsky, Pittel, Dimitrova, and
  Stoitsov]{PhysRevC.65.054319}
Dukelsky, J.; Pittel, S.; Dimitrova, S.S.; Stoitsov, M.V.
\newblock Density matrix renormalization group method and large-scale nuclear
  shell-model calculations.
\newblock {\em Phys. Rev. C} {\bf 2002}, {\em 65},~054319.
\newblock {\url{https://doi.org/10.1103/PhysRevC.65.054319}}.

\bibitem[Pittel and Sandulescu(2006)]{PhysRevC.73.014301}
Pittel, S.; Sandulescu, N.
\newblock Density matrix renormalization group and the nuclear shell model.
\newblock {\em Phys. Rev. C} {\bf 2006}, {\em 73},~014301.
\newblock {\url{https://doi.org/10.1103/PhysRevC.73.014301}}.

\bibitem[Rotureau et~al.(2006)Rotureau, Michel, Nazarewicz, P\l{}oszajczak, and
  Dukelsky]{PhysRevLett.97.110603}
Rotureau, J.; Michel, N.; Nazarewicz, W.; P\l{}oszajczak, M.; Dukelsky, J.
\newblock Density Matrix Renormalization Group Approach for Many-Body Open
  Quantum Systems.
\newblock {\em Phys. Rev. Lett.} {\bf 2006}, {\em 97},~110603.
\newblock {\url{https://doi.org/10.1103/PhysRevLett.97.110603}}.

\bibitem[Tichai et~al.(2023)Tichai, Knecht, Kruppa, Legeza, Moca, Schwenk,
  Werner, and Zarand]{TICHAI2023138139}
Tichai, A.; Knecht, S.; Kruppa, A.; Legeza, {\"O}.; Moca, C.; Schwenk, A.;
  Werner, M.; Zarand, G.
\newblock Combining the in-medium similarity renormalization group with the
  density matrix renormalization group: Shell structure and information
  entropy.
\newblock {\em Phys. Lett. B} {\bf 2023}, {\em 845},~138139.
\newblock
  {\url{https://doi.org/10.1016/j.physletb.2023.138139}}.

\bibitem[Andreozzi et~al.(2003)Andreozzi, Iudice, and Porrino]{Andreozzi_2003}
Andreozzi, F.; Iudice, N.L.; Porrino, A.
\newblock An importance sampling algorithm for generating exact eigenstates of
  the nuclear Hamiltonian.
\newblock {\em J. Phys. G Nucl. Part. Phys.} {\bf 2003},
  {\em 29},~2319.
\newblock {\url{https://doi.org/10.1088/0954-3899/29/10/302}}.

\bibitem[Stumpf et~al.(2016)Stumpf, Braun, and Roth]{PhysRevC.93.021301}
Stumpf, C.; Braun, J.; Roth, R.
\newblock Importance-truncated large-scale shell model.
\newblock {\em Phys. Rev. C} {\bf 2016}, {\em 93},~021301.
\newblock {\url{https://doi.org/10.1103/PhysRevC.93.021301}}.

\bibitem[D et~al.(2010)D, F, N, A, and S]{Bianco_2010}
{Bianco, D.; Andreozzi, F.; Lo Iudice, N.; Porrino, A.; Dimitrova, S.}
\newblock An upgraded version of an importance sampling algorithm for large
  scale shell model calculations.
\newblock {\em J. Phys. Conf. Ser.} {\bf 2010}, {\em
  205},~012002.
\newblock {\url{https://doi.org/10.1088/1742-6596/205/1/012002}}.

\bibitem[Coraggio et~al.(2016)Coraggio, Gargano, and Itaco]{PhysRevC.93.064328}
Coraggio, L.; Gargano, A.; Itaco, N.
\newblock Double-step truncation procedure for large-scale shell-model
  calculations.
\newblock {\em Phys. Rev. C} {\bf 2016}, {\em 93},~064328.
\newblock {\url{https://doi.org/10.1103/PhysRevC.93.064328}}.

\bibitem[Qi(2015)]{Qi849105}
Qi, C.
\newblock Seniority and truncation schemes for the nuclear configuration
  interaction approach.
\newblock {\em Rom. J. Phys.} {\bf 2015}, {\em 60},~782--791.


\bibitem[Poves and Zuker(1981)]{POVES1981235}
Poves, A.; Zuker, A.
\newblock Theoretical spectroscopy and the fp shell.
\newblock {\em Phys. Rep.} {\bf 1981}, {\em 70},~235--314.
\newblock {\url{https://doi.org/10.1016/0370-1573(81)90153-8}}.

\bibitem[Qi et~al.(2016)Qi, Jia, and Fu]{PhysRevC.94.014312}
Qi, C.; Jia, L.Y.; Fu, G.J.
\newblock Large-scale shell-model calculations on the spectroscopy of ${N}<126$
  Pb isotopes.
\newblock {\em Phys. Rev. C} {\bf 2016}, {\em 94},~014312.
\newblock {\url{https://doi.org/10.1103/PhysRevC.94.014312}}.

\bibitem{Wong1}
Wong, S.S.M.
\newblock Truncation of the nuclear shell-model space using spectral strength distribution
\newblock {\em Nucl. Phys. A} {\bf 1978}, {\em 295},~275--288.
\newblock {\url{http://inis.iaea.org/search/search.aspx?orig_q=RN:09377946}}.

\bibitem{Wong2}
Wong, S.S.M.
\newblock {\em Nuclear Statistical Spectroscopy}; Oxford University Press: New York, {NY, USA}, 1986. 

\bibitem{Wong3}
Chang, F.S.; French, J.B.; Thio, T.H.
\newblock Distribution methods for nuclear energies, level densities, and excitation strengths
\newblock {\em Ann. Phys.} {\bf 1971}, {\em 66},~137.
\newblock {\url{https://doi.org/10.1016/0003-4916(71)90186-2}}.

\bibitem[Talmi(1993)]{talmi1993simple}
Talmi, I.
\newblock {\em Simple Models of Complex Nuclei: The Shell Model and Interacting
  Boson Model}; Beitrage zur Wirtschaftsinformatik; Harwood Academic
  Publishers: {Reading, UK}, 1993. 

\bibitem[Qi et~al.(2010)Qi, Wang, Xu, Liotta, Wyss, and Xu]{PhysRevC.82.014304}
Qi, C.; Wang, X.B.; Xu, Z.X.; Liotta, R.J.; Wyss, R.; Xu, F.R.
\newblock Alternate proof of the Rowe-Rosensteel proposition and seniority
  conservation.
\newblock {\em Phys. Rev. C} {\bf 2010}, {\em 82},~014304.
\newblock {\url{https://doi.org/10.1103/PhysRevC.82.014304}}.

\bibitem[Qi(2011)]{PhysRevC.83.014307}
Qi, C.
\newblock Partial conservation of seniority in the $j=9/2$ shell: Analytic and
  numerical studies.
\newblock {\em Phys. Rev. C} {\bf 2011}, {\em 83},~014307.
\newblock {\url{https://doi.org/10.1103/PhysRevC.83.014307}}.

\bibitem[Qi et~al.(2012)Qi, Xu, and Liotta]{QI201221}
Qi, C.; Xu, Z.; Liotta, R.
\newblock Analytic proof of partial conservation of seniority in j = 9/2 shells.
\newblock {\em Nucl. Phys. A} {\bf 2012}, {\em 884--885},~21--35.
\newblock
  {\url{https://doi.org/10.1016/j.nuclphysa.2012.04.007}}.

\bibitem[Xu and Qi(2013)]{XU2013247}
Xu, Z.X.; Qi, C.
\newblock Shell evolution and its indication on the isospin dependence of the
  spin--orbit splitting.
\newblock {\em Phys. Lett. B} {\bf 2013}, {\em 724},~247--252.
\newblock
  {\url{https://doi.org/10.1016/j.physletb.2013.06.018}}.

\bibitem[Jia(2015)]{Jia_2015}
Jia, L.Y.
\newblock Practical calculation scheme for generalized seniority.
\newblock {\em J. Phys. G Nucl. Part. Phys.} {\bf 2015},
  {\em 42},~115105.
\newblock {\url{https://doi.org/10.1088/0954-3899/42/11/115105}}.

\bibitem[Jia and Qi(2016)]{PhysRevC.94.044312}
Jia, L.Y.; Qi, C.
\newblock Generalized-seniority pattern and thermal properties in even Sn
  isotopes.
\newblock {\em Phys. Rev. C} {\bf 2016}, {\em 94},~044312.
\newblock {\url{https://doi.org/10.1103/PhysRevC.94.044312}}.

\bibitem[Maheshwari et~al.(2016)Maheshwari, Jain, and Singh]{MAHESHWARI201662}
Maheshwari, B.; Jain, A.K.; Singh, B.
\newblock Asymmetric behavior of the {B(E2↑;0+ → 2+)} values in 104--130{Sn}
  and generalized seniority.
\newblock {\em Nucl. Phys. A} {\bf 2016}, {\em 952},~62--69.
\newblock
  {\url{https://doi.org/10.1016/j.nuclphysa.2016.04.021}}.

\bibitem[Jiang et~al.(2013)Jiang, Qi, Lei, Liotta, Wyss, and
  Zhao]{PhysRevC.88.044332}
Jiang, H.; Qi, C.; Lei, Y.; Liotta, R.; Wyss, R.; Zhao, Y.M.
\newblock Nucleon pair approximation description of the low-lying structure of
  ${}^{108,109}$Te and ${}^{109}$I.
\newblock {\em Phys. Rev. C} {\bf 2013}, {\em 88},~044332.
\newblock {\url{https://doi.org/10.1103/PhysRevC.88.044332}}.

\bibitem[Jiang et~al.(2014)Jiang, Lei, Qi, Liotta, Wyss, and
  Zhao]{PhysRevC.89.014320}
Jiang, H.; Lei, Y.; Qi, C.; Liotta, R.; Wyss, R.; Zhao, Y.M.
\newblock Magnetic moments of low-lying states in tin isotopes within the
  nucleon-pair approximation.
\newblock {\em Phys. Rev. C} {\bf 2014}, {\em 89},~014320.
\newblock {\url{https://doi.org/10.1103/PhysRevC.89.014320}}.


\bibitem[Qi and Xu(2012)]{PhysRevC.86.044323}
Qi, C.; Xu, Z.X.
\newblock Monopole-optimized effective interaction for tin isotopes.
\newblock {\em Phys. Rev. C} {\bf 2012}, {\em 86},~044323.
\newblock {\url{https://doi.org/10.1103/PhysRevC.86.044323}}.

\bibitem[Machleidt(2001)]{PhysRevC.63.024001}
Machleidt, R.
\newblock High-precision, charge-dependent Bonn nucleon-nucleon potential.
\newblock {\em Phys. Rev. C} {\bf 2001}, {\em 63},~024001.
\newblock {\url{https://doi.org/10.1103/PhysRevC.63.024001}}.

\bibitem[Hjorth-Jensen et~al.(1995)Hjorth-Jensen, Kuo, and
  Osnes]{HJORTHJENSEN1995125}
Hjorth-Jensen, M.; Kuo, T.T.; Osnes, E.
\newblock Realistic effective interactions for nuclear systems.
\newblock {\em Phys. Rep.} {\bf 1995}, {\em 261},~125--270.
\newblock {\url{https://doi.org/10.1016/0370-1573(95)00012-6}}.

\bibitem[Sn()]{Sn}
 Available online: \url{http://www.nuclear.kth.se/cqi/sn100/}  ({14 April 2024}).
 
\bibitem{PhysRevLett.99.162501}
Vaman, C.; Andreoiu, C.; Bazin, D.; Becerril, A.; Brown, B.A.; Campbell, C.M.; Chester, A.; Cook, J.M.; Dinca, D.C.; Gade, A.; et al.
\newblock $Z=50$ Shell Gap near $^{100}\mathrm{Sn}$ from Intermediate-Energy Coulomb Excitations in Even-Mass $^{106\ensuremath{-}112}\mathrm{Sn}$ Isotopes.
\newblock {\em Phys. Rev. Lett.} {\bf 2007}, {\em 99},~162501.
\newblock {\url{https://doi.org/10.1103/PhysRevLett.99.162501}}.

 
\bibitem{PhysRevC.72.061305}
Banu, A.; Gerl, J.; Fahlander, C.; G\'orska, M.; Grawe, H.; Saito, T.R.; Wollersheim, H.-J.; Caurier, E.; Engeland, T.; Gniady, A.; et al.
\newblock $^{108}\mathrm{Sn}$ studied with intermediate-energy Coulomb excitation.
\newblock {\em Phys. Rev. C} {\bf 2005}, {\em 72},~061305.
\newblock {\url{https://doi.org/10.1103/PhysRevC.72.061305}}.

\bibitem[Möller et~al.(2019)Möller, Mumpower, Kawano, and Myers]{MOLLER20191}
Möller, P.; Mumpower, M.; Kawano, T.; Myers, W.
\newblock Nuclear properties for astrophysical and radioactive-ion-beam
  applications (II).
\newblock {\em At. Data Nucl. Data Tables} {\bf 2019}, {\em
  125},~1--192.
\newblock {\url{https://doi.org/10.1016/j.adt.2018.03.003}}.

\bibitem[NND()]{NNDC}
Data Extracted Using the {NNDC} World Wide Web Site from the ENSDF.
 Available online: \url{https://www.nndc.bnl.gov/ensdf/} ({10 Sept. 2024}).

\bibitem[Sahoo et~al.(2024)Sahoo, Srivastava, Shimizu, and
  Utsuno]{PhysRevC.110.024306}
Sahoo, S.; Srivastava, P.C.; Shimizu, N.; Utsuno, Y.
\newblock Nuclear structure properties of $^{193-200}\mathrm{Hg}$ isotopes
  within large-scale shell model calculations.
\newblock {\em Phys. Rev. C} {\bf 2024}, {\em 110},~024306.
\newblock {\url{https://doi.org/10.1103/PhysRevC.110.024306}}.

\bibitem[Sandzelius et~al.(2007)Sandzelius, Hadinia, Cederwall, Andgren,
  Ganio\ifmmode~\breve{g}\else \u{g}\fi{}lu, Darby, Dimmock, Eeckhaudt, Grahn,
  Greenlees, Ideguchi, Jones, Joss, Julin, Juutinen, Khaplanov, Leino, Nelson,
  Niikura, Nyman, Page, Pakarinen, Paul, Petri, Rahkila, Sar\'en, Scholey,
  Sorri, Uusitalo, Wadsworth, and Wyss]{PhysRevLett.99.022501}
Sandzelius, M.; Hadinia, B.; Cederwall, B.; Andgren, K.;
  Ganio\ifmmode~\breve{g}\else \u{g}\fi{}lu, E.; Darby, I.G.; Dimmock, M.R.;
  Eeckhaudt, S.; Grahn, T.; Greenlees, P.T.;  et~al.
\newblock Identification of Excited States in the ${T}_{z}=1$ Nucleus
  $^{110}\mathrm{Xe}$: Evidence for Enhanced Collectivity near the $N=Z=50$
  Double Shell Closure.
\newblock {\em Phys. Rev. Lett.} {\bf 2007}, {\em 99},~022501.
\newblock {\url{https://doi.org/10.1103/PhysRevLett.99.022501}}.
  
\end{thebibliography}
\end{document}